\documentstyle[12pt]{article}
\input epsf.sty
\def\ZZZ{{\hbox{ Z\kern-1.6mm Z}}}

\newcommand{\eps}{\epsilon}
\newcommand{\ra}{\rangle}
\newcommand{\ov}{\overline}
\newcommand{\la}{\langle}

\newcommand{\vp}{{\large \varphi}}

\newcommand{\tl}{\lambda}

\newcommand{\KK}{{\cal K}}
\newcommand{\BB}{{\cal B}}

\newcommand{\GG}{{\cal G}}

\newcommand{\FF}{{\cal F}}

\newcommand{\OO}{{\cal O}}
\newcommand{\QQ}{{\cal Q}}
\newcommand{\PP}{{\cal P}}

\newcommand{\wt}{\widetilde}
\newcommand{\wh}{\widehat}
\newcommand{\wc}{\check}

\newcommand{\RR}{{\cal R}}

\newcommand{\SSS}{{\cal S}}

\newcommand{\be}{\begin{equation}}
\newcommand{\ee}{\end{equation}}
\newcommand{\ben}{\begin{eqnarray}\displaystyle}
\newcommand{\een}{\end{eqnarray}}
\newcommand{\refb}[1]{(\ref{#1})}
\newcommand{\p}{\partial}
\newcommand{\sectiono}[1]{\section{#1}\setcounter{equation}{0}}

\def\one{{\hbox{ 1\kern-.8mm l}}}
\def\zero{{\hbox{ 0\kern-1.5mm 0}}}

\begin{document}
{}~
{}~
\hfill\vbox{\hbox{hep-th/0408064}%\hbox{MRI-P-020401}
}\break

\vskip .6cm

\centerline{\Large \bf Symmetries, Conserved Charges and (Black)
Holes}
\medskip
\centerline{\Large \bf
in Two Dimensional String Theory}

\medskip

\vspace*{4.0ex}

\centerline{\large \rm
Ashoke Sen}

\vspace*{4.0ex}

\centerline{\large \it Harish-Chandra Research
Institute}

\centerline{\large \it  Chhatnag Road, Jhusi,
Allahabad 211019, INDIA}

\medskip

\centerline{E-mail: ashoke.sen@cern.ch,
sen@mri.ernet.in}

\vspace*{5.0ex}

\centerline{\bf Abstract} \bigskip

Two dimensional string theory is known to have an infinite
dimensional symmetry, both in the continuum formalism as well as
in the matrix model formalism. We develop a systematic procedure
for computing the conserved charges associated with these
symmetries for any configuration of D-branes in the continuum
description. We express these conserved charges in terms of the
boundary state associated with the D-brane, and also in terms of
the asymptotic field configurations produced by this D-brane.
Comparison of the conserved charges computed in the continuum
description with those computed in the matrix model description
facilitates identification of the states between these two
formalisms.  Using this we put constraints on the continuum
description of the hole states in the matrix model, and matrix
model description of the black holes solutions of the continuum
theory. We also discuss possible generalization of the
construction of the conserved charges to the case of D-branes in
critical string theory.

\vfill \eject

\baselineskip=16pt

\tableofcontents

\sectiono{Introduction and Summary} \label{s1}

Recent investigation in two dimensional string
theory\cite{0304224,0305159,0305194,0307083,0307195} has shown
that they can provide us with a useful arena for studying various
general properties of string theory, most notably the relationship
between the open and closed string description of unstable D-brane
systems\cite{0308068,0305011,0306137,0312135,0312163, 0312192}.
(See \cite{0312196,0406106} for aspects of open-closed string
duality for stable D-branes in this string theory.). The feature
that makes this theory most useful is that it has two different
formulations. The first one, known as the continuum
description\cite{DAVID,DK} (see also \cite{KPZ}), follows the
usual formulation of string theory based on a world-sheet action
containing a matter part with central charge 26 and a ghost part
with central charge $-26$. The matter part in turn consists of a
free time-like scalar field $X^0$ of central charge 1 and a
Liouville scalar field $\vp$ with an exponentially growing
potential. A linear dilaton background along the Liouville
direction makes the Liouville theory have total central charge 25.
In this formalism the theory can be studied using the usual string
perturbation theory based on genus expansion. The other
formulation of the theory, known as the matrix model, is based on
discretizing the world sheet path integral and taking an
appropriate double scaling limit\cite{GROMIL,BKZ,GINZIN}. This in
turn can be shown to be equivalent to a theory of free
non-interacting fermions, each moving under an inverted harmonic
oscillator potential. The vacuum of the theory is a state in which
all levels below a certain fixed energy are filled and all level
above this energy are empty. In this formulation we can easily
analyze the system to all orders in perturbation theory.

The usual closed string states of the continuum string theory are
related to the matrix model states by bosonization of the fermion
field followed by a non-local field
redefinition\cite{DASJEV,SENWAD,GROKLE}. While early work on this
subject focussed on the comparison of the properties of closed
strings in the two formulations, the recent surge of interest in
this subject arises from the study of D-branes in the two
descriptions of the theory. The continuum version of the theory
admits an unstable D0-brane configuration with an open string
tachyon on its world-volume\cite{0101152}. Following the general
method developed in \cite{0203211,0203265} one can construct an
exact classical solution describing the open string tachyon
rolling away from the maximum of the potential. By studying the
closed string description of this process in the continuum string
theory following \cite{0303139,0304192,0306132}, and comparing
this with the single fermion excitation in the matrix model using
the known relation between the states of the matrix model and the
closed string states in the continuum description, it was
concluded in \cite{0305159} that the rolling tachyon configuration
on a single D0-brane in the continuum theory describes precisely
single fermion excitations in the matrix model.

Despite this new understanding of the relationship between the
matrix model and continuum description of two dimensional string
theories, several questions remain unanswered. In particular we
still do not have a complete map between the known states of the
continuum theory and known states of the matrix model. For example
the matrix model, besides containing fermionic excitations, also
contains hole like excitations where we remove a fermion from an
energy level below the fermi level. A completely convincing
description of these states in the continuum theory is still
missing (although some candidates have been proposed in
\cite{0307195,0307221}). On the other hand the continuum version
of this theory admits black hole solutions\cite{MSW,WB}. Although
there are some proposals for a representation of the Euclidean
black hole in the matrix model\cite{0101011,0311177}, a satisfactory
description of these black hole states in the Lorenzian version of
the matrix model is still lacking.

Both the continuum version and the matrix model version of the
theory are known to have infinite number of global
symmetries\cite{SENWAD,MOORESEI,MINIC,9108004,9109032,9110021,9209036,
9210105,9302106,9201056,9507041}, and hence associated with them
there must be infinite number of conserved charges. Thus if we can
find the precise relation between the conserved charges in the
continuum description and those in the matrix model description,
then comparison of these conserved charges could provide us useful
guidelines for making suitable identification between the states
in the continuum description and states in the matrix model
description. The construction of the conserved charges in the
matrix model description follows straightforward application of
Noether's method. Thus the main issue is to construct these
conserved charges in the continuum description and relate them to
the conserved charges in the matrix model description. This
program was initiated in a previous paper\cite{0402157} where it
was shown that requirement of BRST invariance constrains the time
evolution of certain components of the boundary state describing a
D-brane, and hence leads to certain conserved charges. By
evaluating these conserved charges for the rolling tachyon
configuration, and comparing them with the conserved charges of
the same configuration in the matrix model description, we found
the relationship between these two sets of conserved charges.
However in this analysis the construction of the conserved charges
on the continuum side was somewhat ad hoc, in the sense that it
required assuming a certain structure of the boundary state, and
within that structure requirement of BRST invariance fixed the
time dependence of certain terms of the boundary state. A general
procedure for constructing the conserved charges for a general
boundary state was not given. Also the conserved charges
constructed this way did not get related to any specific symmetry
of the theory.

The main goal of this paper will be to develop a systematic method
for computing the charge carried by a D-brane associated with a
given a global symmetry transformation in any string theory, and
then apply this to two dimensional string theory. In particular,
we derive a general expression for the conserved charges carried
by any D-brane system in terms of its boundary state. Comparison
of the charges carried by a D0-brane in the matrix model and the
continuum description of the two dimensional string theory then
allows us to find a precise relation between the conserved charges
in the two descriptions. Using these relations we can put
constraints on what kind of D-brane describes the hole states of
the matrix model, and also what kind of matrix model configuration
describes the black hole states of the continuum string theory.

The paper is organized as follows. In section \ref{s2} we review
how a rigid closed string gauge transformation, for which the
field independent part of the transformation vanishes, generates
global symmetries of the open string field theory\cite{9705241}.
Associated with this global symmetry we can associate a conserved
charge. We develop an algorithm for computing this conserved
charge for any D-brane system in terms of the boundary state
describing the D-brane. The final formula for the conserved charge
is given in eq.\refb{e9}. The simplest global symmetry of this
kind is time translation, and the associated conserved charge
gives the energy of the D-brane.

In section \ref{s3a} we review the structure of rigid closed
string gauge transformations in the two dimensional string theory.
These are obtained from the elements of relative BRST cohomology
in the ghost number one sector, and have been classified in
refs.\cite{LIAN,9108004,MMS1,BMP,9201056}. There are infinite
number of such rigid gauge transformations, labelled by SU(2)
quantum numbers $(j,m)$ with $j\ge 1$, $m=-(j-1), -(j-2), \ldots
(j-1)$. Thus associated with these transformations there are
infinite number of conserved charges $Q_{j,m}$. Using the result
of section \ref{s2} we explicitly write down in eqs.\refb{e20},
\refb{edefqjm} the expression for $Q_{j,m}$ carried by a D-brane
of two dimensional string theory in terms of the boundary state of
the D-brane.

In section \ref{s3b} we evaluate these conserved charges for a
specific D-brane system, -- namely the rolling tachyon
configuration on the unstable D0-brane of two dimensional string
theory. These configurations are labelled by a single parameter
$\tl$, and we find in eq.\refb{enewqjm} explicit expression for
$Q_{j,m}$ as a function of the parameter $\tl$.

In \cite{0402157} we had calculated the discrete state closed
string background produced by the rolling tachyon configuration in
the weak coupling region of large negative $\vp$. We recall these
results in section \ref{sasrol} (eq.\refb{eback1}), and point out
that there are some additional contributions to this closed string
background due to some subtleties which were overlooked in
\cite{0402157}. Eq.\refb{ere6} gives a typical example of such
additional contributions.

The matrix model description of the theory also contains an
infinite set of conserved charges $W_{l,m}$ with $2m\in Z$,
$(l-m)\in Z$, $l\ge |m|$. In section \ref{s4} we compare the
conserved charges of the matrix model description with the
conserved charges of the continuum string theory. By evaluating
the conserved charges $W_{l,m}$ of single fermion excitations in
the matrix model, and comparing the $\tl$ and time dependence of
these charges with those of $Q_{j,m}$ carried by a single D0-brane
in the continuum description, we determine in eq.\refb{e37a} the
relation between the conserved charges of the matrix model and
those in the continuum description of the theory. Although these
relations are derived by evaluating the charges in the two
descriptions of the D0-brane, once the relationship is determined,
it must hold for any other system.

In section \ref{s5} we try to find a continuum description of the
hole states of the matrix model with the help of the conserved
charges of the theory. From the description of the conserved
charges in the fermionic formulation of the matrix model it is
easy to evaluate these conserved charges for a given hole state.
These are given in eq.\refb{ewhole}.  Using the known relation
between the conserved charges in the matrix model and those
($Q_{j,m}$'s) in the continuum theory, we calculate in
eq.\refb{e37b} the expected $Q_{j,m}$ for the hole states.
Whatever be the continuum description of the hole must carry the
same charges. On the other hand our analysis of section \ref{s3a}
gives an explicit expression for the $Q_{j,m}$'s in terms of the
boundary state of the D-brane. This gives constraints on the
boundary state describing a hole state, but does not determine it
completely. The proposals made in \cite{0307195,0307221} satisfy
these constraints; however we argue that they do not reproduce the
expected closed string profile of a hole state. We propose a
possible candidate for these hols states based on some qualitative
arguments, but there is no definite conclusion yet.

Earlier work on the relation between symmetries of the matrix
model and those in the continuum description was carried out in
the limit of zero potential for the Liouville field, and was based
on the comparison of the symmetry algebras in the two
descriptions. In order to compare our results to the earlier
results, we study in section \ref{smu} the limit of our results as
we take the potential for the Liouville field to zero. We show
first of all that this limit exists and gives a well defined
relation \refb{emore3} between the conserved charges in the
continuum theory and those in the matrix model. Furthermore, up to
normalization factors these relations are consistent with the
earlier results obtained by working directly with the Liouville
theory without any potential term. We also study the limit of the
discrete state closed string background produced by the D0-brane
boundary state as we take the Liouville potential to zero keeping
the energy of the D0-brane fixed. We find that the closed string
field configuration approaches a finite value \refb{eback9} in
this limit.

The analysis of section \ref{s3a} expresses the conserved charges
carried by a D-brane in terms of its boundary state. This relation
is specific to D-branes since only D-branes are described by
boundary states. In section \ref{sadm} we manipulate the results
of section \ref{s3a} to rewrite the conserved charge in terms of
the asymptotic values of certain components of closed string
fields produced by the brane for large negative $\vp$. The final
formula, given in \refb{eadm19}, is the analog of the Gauss law
for electrodynamics relating the electric charge to asymptotic
electric field, or the ADM formula in gravity relating the mass of
a system to the asymptotic gravitational field. These relations
between conserved charges and asymptotic closed string fields are
expected to hold for any system in two dimensional string theory,
even if the system cannot be regarded as a collection of D-branes.

In section \ref{sblack} we use this result to determine the
conserved charges carried by the black hole solution of the two
dimensional string theory. For simplicity the analysis of this
section is carried out for vanishing potential for the Liouville
field, which in the matrix model description corresponds to the
fermi level coinciding with the maximum of the potential. Although
the black hole was initially constructed as a solution of the
effective field theory\cite{MSW} or as an exact conformal field
theory\cite{WB}, it is possible to represent it as a solution in
string field theory by using an iterative procedure for solving
the equations of motion of string field theory\cite{MMS}. Using
this we can find the asymptotic closed string field configuration
associated with the black hole and hence the charges $Q_{j,m}$
carried by the black hole. We find that the black hole carries
only the conserved charge $Q_{1,0}$; all other charges $Q_{j,m}$
for $(j,m)\ne (1,0)$ vanish. Using the known relation \refb{e37a}
between $Q_{j,m}$ and the conserved charges in the matrix model
description of the system we can then constrain the possible
configurations in the matrix model which are compatible with these
conserved charges. In particular we find that the matrix model
description of the black hole must consist of a large number of
low energy fermion-hole pairs instead of a finite number of finite
energy fermions and holes.

This description of the black hole poses an apparent puzzle. Since
in the matrix model description the fermions are non-interacting,
and since the black hole background differs from the usual vacuum
in terms of creation of a large number of low energy fermion-hole
pairs, a classical D0-brane carrying finite energy should not be
able to recognize the difference between a black hole and the
ordinary vacuum. Can this be true in two dimensional string
theory? While a complete answer to this question requires studying
the D0-brane motion in these backgrounds to all orders in
$\alpha'$, we show that at least in the approximation where we
take the D0-brane world-line action to be of the Dirac-Born-Infeld
form, the classical D0-brane cannot distinguish the black hole
from the usual linear dilaton background. This is shown by
demonstrating that there is a coordinate transformation that
converts the effective metric seen by the D0-brane in the black
hole background to the effective metric seen by the D0-brane to
the linear dilaton background. This coordinate transformation acts
only on the space coordinate and leaves the time coordinate
unchanged. This is consistent with the fact that both for the
black hole and the usual flat background with a linear dilaton
field, the time coordinate should be identified with the time
coordinate of the matrix model.

Although the main emphasis of the paper has been on the
construction of the conserved charges and their interpretation in
the two dimensional string theory, we can try to generalize the
construction to critical string theory by replacing the primary
vertex operators in the Liouville field theory by appropriate
primary vertex operators in the critical string theory, and by
replacing the Liouville Virasoro generators by the total Virasoro
generators associated with all the 25 space-like coordinate fields
in the critical string theory. There are however various subtle
issues in this approach. There are discussed in section
\ref{scrit}.

Finally the appendices contain some technical results which are
required for the explicit construction of conserved charges and their 
normalization in two
dimensional string theory. 

\sectiono{Symmetries to Conserved Charges in Open String Theory} \label{s2}

In this section we briefly outline the general procedure for
obtaining the conserved charge in classical open string theory
associated with a specific global symmetry. We shall focus on the
global symmetries associated with rigid gauge transformations in
closed string theory\cite{9705241}.  An example of this is
space-time translation symmetry, which can be thought of as a
rigid general coordinate transformation.

We shall carry out the discussion in the context of string field
theory. We begin with some version of covariant open-closed string
field theory\cite{9705241} formulated for a given D-brane in a
given space-time background. However our analysis will be quite
general and we shall not restrict ourselves to any specific form
of the action. Let us denote by $\{\Phi_\alpha\}$ the closed
string degrees of freedom and by $\{\Psi_r\}$ the open string
degrees of freedom, with the indices $\alpha$ and $r$, besides
containing discrete labels, also carrying information about
momenta of the fields along non-compact space-time directions.
Then the open-closed string field theory action has the
form\cite{9705241}:
 \be
\label{eop1}
{1\over g_s^2} S_0(\Phi) + {1\over g_s} S_1(\Phi,
\Psi) + \OO(g_s^0)\, ,
 \ee
where $g_s$ denotes string coupling constant. The order $g_s^{-2}$
and $g_s^{-1}$ terms get contributions respectively from the
sphere and disk correlation functions of the world-sheet theory.
Let $D$ denote the dimension of space-time in which the closed
string theory lives. Then a typical closed string gauge
transformation is parametrized by some arbitrary function
$\epsilon(p)$ of $D$ dimensional momentum $p$. The infinitesimal
gauge transformation laws take the form:
 \ben \label{eop2}
  \delta\Phi_\alpha &=& \sum_{n=0}^\infty g_s^n \,
  \delta\Phi^{(n)}_\alpha = \int d^D p \, \, \eps(p) \, \left[
  h_\alpha^{(0)}(\Phi, p) + g_s h_\alpha^{(1)}(\Phi, \Psi, p) +
  \OO(g_s^2)\right]\nonumber \\
 \delta \Psi_r &=& \sum_{n=0}^\infty g_s^n \,
  \delta\Psi^{(n)}_r = \int d^D p \, \, \eps(p) \, \left[
f_r^{(0)}(\Phi,\Psi, p) + \OO(g_s)\right]\, ,
 \een
for suitable function $h_\alpha^{(n)}$ and $f_r^{(n)}$. The
contributions to $h^{(0)}_\alpha$ come from sphere correlation
functions, whereas the contributions to $h^{(1)}_\alpha$ and
$f^{(0)}_r$ come from disk correlation functions. Note that the
leading contribution to the action and the leading contribution to
$\delta\Phi_\alpha$ do not depend on the open string fields
$\Psi_r$. We call this action and gauge transformation laws tree
level closed string action and gauge transformation laws
respectively. On the other hand
 \be \label{eop3}
S_{open} (\Psi) \equiv {1\over g_s} S_1(\Phi=0, \Psi)
 \ee
 is called the tree level open string field theory action
in the $\Phi=0$ closed string background.\footnote{A family of
open closed string field theory was constructed in \cite{9705241},
and Witten's open string field theory\cite{osft} appears as the
open string sector of a special member of this family.} Invariance
of the full action \refb{eop1} under the gauge transformation laws
\refb{eop2} gives:
 \be \label{eop4}
 {\delta S_0(\Phi) \over \delta \Phi_\alpha} \delta \Phi^{(0)}_\alpha =
 0\, ,
 \ee
 \be \label{eop5}
{\delta S_1(\Phi, \Psi) \over \delta \Phi_\alpha} \delta
\Phi^{(0)}_\alpha + {\delta S_1(\Phi, \Psi) \over \delta \Psi_r}
\delta \Psi^{(0)}_r +{\delta S_0(\Phi) \over \delta \Phi_\alpha}
\delta \Phi^{(1)}_\alpha = 0\,
 \ee
etc. We shall choose the string field variables such that $\Phi=0$
is trivially a solution of the classical closed string field
equations. Thus $\delta S_0/\delta \Phi_\alpha$ vanishes at
$\Phi=0$. Putting $\Phi=0$ in eq.\refb{eop5} and using
\refb{eop2}, \refb{eop3} we get
 \be \label{eop6}
\left[{\delta S_1(\Phi, \Psi) \over \delta
\Phi_\alpha}\bigg|_{\Phi=0} \,  h_\alpha^{(0)}(\Phi=0, p) +
{\delta S_{open}(\Psi) \over \delta \Psi_r} \, f_r^{(0)}(\Phi=0,
\Psi, p) \right] =0\, .
 \ee

In general $h_\alpha^{(0)}(\Phi=0, p)$ is non-zero. However
suppose for some special value of the momentum $p$ it vanishes:
 \be \label{eop7}
 h^{(0)}_\alpha(\Phi=0, p=c) = 0\, .
 \ee
 Physically it means that the field independent
 term in the tree level closed string gauge transformation law
 vanishes. In other words we have a rigid gauge transformation
 that leaves the $\Phi=0$ background unchanged.
 Putting $p=c$ in \refb{eop6} we now get
 \be \label{eop8}
{\delta S_{open}(\Psi) \over \delta \Psi_r} \, f_r^{(0)}(\Phi=0,
\Psi, p=c) = 0\, .
 \ee
This describes a global symmetry of the tree level open string
field theory with infinitesimal transformation law
 \be \label{einfop}
 \delta \Psi_r = \eps \, f^{(0)}_r(\Phi=0, \Psi, p=c)\, .
  \ee
   Thus we see
that associated with every rigid closed string gauge
transformation that leaves the $\Phi=0$ background unchanged, we
have a global symmetry of the classical open string field
theory\cite{9705241}.

Given a global symmetry, there should be a conserved charge
associated with this symmetry. One possible way to find this
charge will be to use Noether method. The symmetry transformations
are non-local, but a general Noether method for finding the
conserved charge associated with a non-local symmetry
transformation in a non-local theory was outlined in
\cite{0403200}. We can follow the same method for finding the
expression for the conserved charge associated with these global
symmetries of open string theory. However this procedure only
gives the difference between the conserved charge carried by a
given open string field configuration, and that carried by the
$\Psi=0$ configuration representing the original D-brane on which
we have formulated the open string field theory. In particular if
we set the open string field $\Psi$ to zero, then the expression
for the conserved charge vanishes. Our main interest on the other
hand will be in the expression for the conserved charge that the
original D-brane carries. For this we need to use a different
method which we shall describe now.

The basic procedure can be understood in analogy with the
computation of the energy momentum tensor of a D-brane. Defining
the energy-momentum tensor in the open string field theory through
the Noether prescription gives correctly the difference in the
energy-momentum tensor between two open string
configurations\cite{0403200}, but this does not give the
energy-momentum tensor of the D-brane itself. The latter can be
calculated by examining the coupling of the  metric to the D-brane
world-volume. In a similar spirit one would expect that the
information about all other conserved charges carried by the
D-brane, which are associated with rigid gauge transformations
that leave the closed string background $\Phi=0$ invariant, should
also be calculable by examining the coupling of the various closed
string modes to the D-brane. In fact the relevant information is
already contained in eq.\refb{eop6}.
 Using eq.\refb{eop7} and assuming that $h^{(0)}_\alpha(\Phi=0, p)$ is
analytic at $p=c$, we can write
 \be \label{en4}
h^{(0)}_\alpha(\Phi =0, p) = (p_\mu - c_\mu) \, \wh
h_\alpha^\mu(p) \, ,
 \ee
for some $\wh h_\alpha^\mu$. If we define
 \be \label{eop10}
 \GG_\alpha(\Psi) = {\delta S_1(\Phi, \Psi)\over
 \delta\Phi_\alpha}\bigg|_{\Phi=0}\, ,
 \ee
 then eq.\refb{eop6} can be rewritten as
 \be \label{en5}
(p_\mu-c_\mu) \, \, \GG_\alpha(\Psi) \, \, \wh h_\alpha^\mu(p) +
 {\delta S_{open}(\Psi) \over  \delta\Psi_r}\, f_r^{(0)}(\Phi=0,
 \Psi, p) =0 \, .
 \ee
Now if the fields $\Psi_s$ satisfy their equations of motion then
$\delta S_{open}(\Psi) /\delta\Psi_r=0$. In this case we have
 \be \label{en7}
(p_\mu-c_\mu) \, \GG_\alpha(\Psi) \, \wh h_\alpha^\mu(p) = 0\, .
 \ee
If we define
 \be \label{en8}
F^\mu(x) = \int d^D p \, e^{-ip.x} \, \GG_\alpha(\Psi) \wh
h_\alpha^\mu(p)
 \ee
then \refb{en7} may be rewritten as
 \be \label{en10}
\p_\mu \left(e^{ic.x} F^\mu(x)\right) =0\, .
 \ee
This gives the conserved charge
 \be \label{en11}
\int d^{D-1} x \,  e^{ic.x} F^0(x)\, .
 \ee

We shall now make this construction more explicit by working with
specific representation of closed string fields. We shall restrict
our analysis to a closed string background in which the time
direction has associated with it a world-sheet conformal field
theory (CFT) of a free scalar field $X^0$ which does not couple to
any other world-sheet field. In this case we can work with the
Euclidean continuation of the theory obtained by the replacement
$x^0\to -i x$. We can represent the closed string field by a state
$|\Phi\ra$ of ghost number two in the bulk CFT on a cylinder,
satisfying
 \be
\label{e1} (b_0-\bar b_0) |\Phi\ra = 0, \qquad (L_0-\bar L_0)
|\Phi\ra = 0\, ,
 \ee
where $b_n$, $\bar b_n$, $c_n$, $\bar c_n$ denote the usual ghost
oscillators, and $L_n$, $\bar L_n$ denote the total Virasoro
generators of the world-sheet theory of matter and ghost fields.
Closed string gauge transformations in this theory are generated
by ghost number one states $|\Lambda\ra$ of the CFT on a cylinder,
satisfying
 \be \label{e2} (b_0-\bar b_0) |\Lambda\ra = 0,
\qquad (L_0-\bar L_0) |\Lambda\ra = 0\, .
 \ee
The effect of the infinitesimal gauge transformations on the
closed string fields is given by:
 \be \label{e3} \delta |\Phi\ra = (Q_B+\bar Q_B)|\Lambda\ra +\OO(\Phi)\, ,
 \ee
where $Q_B$ and $\bar Q_B$ are the holomorphic and antiholomorphic
components of the BRST charge in closed string theory. Thus for a
$|\Lambda\ra$ satisfying
 \be \label{e4}
(Q_B+\bar Q_B)|\Lambda\ra = 0\, ,
 \ee
the infinitesimal gauge transformation of $|\Phi\ra$ vanishes at
$|\Phi\ra=0$. As a result this gauge transformation leaves the
$|\Phi\ra=0$ background unchanged. By our previous argument, this
must generate a symmetry of the pure open string field theory
living on a D-brane, and give rise to a conserved charge in this
theory.\footnote{Of course we can trivially find solutions to
\refb{e4} by choosing $|\Lambda\ra=(Q_B+\bar Q_B) |s\ra$ for some
ghost number zero state $|s\ra$ satisfying conditions similar to
those in \refb{e1}, \refb{e2}. But one can show that these
generate gauge symmetries in the open string field theory, and do
not give rise to any additional global symmetry\cite{9705241}. We
shall see later that in this case the associated conserved charge
vanishes identically.}

Our goal will be to construct expressions for these conserved
charges explicitly for any D-brane system living in this closed
string background. For simplicity we shall evaluate the charge in
trivial open string background $\Psi=0$, -- this will evaluate the
charge carried by the specific D-brane used in the construction of
the open closed string field theory without any further open
string excitations on the brane. Let $|\Lambda(p)\ra$ denote a
family of closed string gauge transformation parameters labelled
by $X$ momentum $p$, such that $|\Lambda(p=c)\ra=|\Lambda\ra$ for
some special momentum $c$.\footnote{We are assuming that the other
momentum components have already been set equal to the specific
values for which $(Q_B+\bar Q_B)|\Lambda(p)\ra$ vanishes.} Then
$(Q_B+\bar Q_B)|\Lambda(p)\ra$ vanishes at $p=c$ and we can write
 \be \label{e5}
(Q_B+\bar Q_B)|\Lambda(p)\ra = (p-c) |\phi(p)\ra\, ,
 \ee
where $|\phi(p)\ra$ is some ghost number two state. Now since
$(Q_B+\bar Q_B)$ is nilpotent, we see from eq.\refb{e5} that
$|\phi(p)\ra$ is BRST invariant for all $p\neq c$, and hence by
analytic continuation BRST invariant also for $p=c$. Furthermore
it has the property that for any $p\ne c$ it is BRST trivial, but
for $p=c$ it could be a non-trivial element of the BRST cohomology
in the ghost number two sector. We shall see later that we can get
non-trivial conserved charges only if $|\phi(p=c)\ra$ is not BRST
trivial.

Let us denote by $|\BB\ra$ the boundary state associated with the D-brane
on which we have formulated the open string theory. Then the full string
field theory action contains a coupling:
 \be \label{e6}
\la \BB | (c_0-\bar c_0) |\Phi\ra \, .
 \ee
Invariance of this term under the infinitesimal gauge
transformation \refb{e3} generated by the family of gauge
transformation parameters $|\Lambda(p)\ra$ requires:
 \be \label{e7}
\la \BB| (c_0- \bar c_0) \,(Q_B+\bar Q_B)  |\Lambda(p)\ra = 0\, ,
 \ee
 For ordinary D-branes eq.\refb{e7} follows from the BRST
 invariance of $\la \BB|$ and the analogs of \refb{e1}, \refb{e2}:
 \be \label{ebcons}
 \la \BB|(Q_B+\bar Q_B)=0, \quad \la \BB| (b_0-\bar b_0)=0, \quad \la
 \BB|(L_0-\bar L_0)=0\, .
 \ee
 This allows us to replace $(c_0-\bar c_0)(Q_B+\bar Q_B)$ by
 $\{(c_0-\bar c_0),(Q_B+\bar Q_B)\}$ in \refb{e7}. This does not
 have any zero mode of  $(c_0-\bar c_0)$ and hence the matrix
 element vanishes.

 Using \refb{e5}, eq.\refb{e7} becomes:
 \be \label{e8}
(p-c) \la \BB| (c_0- \bar c_0) |\phi(p)\ra = 0\, .
 \ee
If we define:
 \be \label{e9}
F(x) = \int {dp\over 2\pi} \, e^{-ip x} \la \BB| (c_0- \bar c_0)
|\phi(p) \ra \, ,
 \ee
then \refb{e8} may be rewritten as
 \be \label{e11} \p_x\left( e^{ic x}
F(x)\right) = 0\, .
 \ee
Replacing $x$ by $ix^0$ we now get:
 \be
\label{e11a} \p_0 \left( e^{-c x^0} F(i x^0)\right) = 0\, .
 \ee
Thus
$e^{-c x^0} F(ix^0)$ is a conserved charge. This gives a general
procedure for constructing the conserved charge carried by a D-brane
corresponding to a specific rigid gauge transformation in closed string
theory. The suggestion that the BRST invariance of $\la\BB|$ carries
information about conserved charges has been made earlier in
\cite{0309074}.

Given an element $|\Lambda\ra$ of the BRST cohomology with a
specific momentum $c$, there are clearly infinite number of
families $|\Lambda(p)\ra$ with the property that
$|\Lambda(p=c)\ra=|\Lambda\ra$. On physical grounds the conserved
charge associated with the symmetry generated by $|\Lambda\ra$
should not depend on the choice of the family. We shall now prove
this explicitly by demonstrating that if two families of gauge
transformations parameters $|\Lambda^{(1)}(p)\ra$ and
$|\Lambda^{(2)}(p)\ra$ approach the same value at $p=c$, then they
give rise to the same conserved charge. In this case, we may write
 \be \label{ediff1}
|\Lambda^{(1)}(p)\ra - |\Lambda^{(2)}(p)\ra = (p-c)
|\Lambda^{(0)}(p)\ra\, ,
 \ee
for some $|\Lambda^{(0)}(p)\ra$, so that the difference between
$|\Lambda^{(1)}(p)\ra$ and $|\Lambda^{(2)}(p)\ra$ vanishes at
$p=c$. Eqs.\refb{e5} and \refb{ediff1} now give:
 \be \label{ediff2}
|\phi^{(1)}(p)\ra - |\phi^{(2)}(p)\ra  = (Q_B+\bar Q_B)
|\Lambda^{(0)}(p)\ra  \, ,
 \ee
where $|\phi^{(i)}(p)\ra$ is related to $|\Lambda^{(i)}(p)\ra$ as
in eq.\refb{e5}. If $F^{(1)}(x)$ and $F^{(2)}(x)$ denote the
corresponding conserved charges as defined in \refb{e9}, then we
have
 \be \label{ediff3}
F^{(1)}(x) - F^{(2)}(x) =\int {dp\over 2\pi} \, e^{-ip x}  \la
\BB|(c_0-\bar c_0) (Q_B+\bar Q_B) |\Lambda^{(0)}(p)\ra   \, .
 \ee
Since $\la \BB|$ is annihilated by $(Q_B+\bar Q_B)$, we can
replace the $(c_0-c_0^-) (Q_B+\bar Q_B)$ term by $\{
(c_0-c_0^-), (Q_B+\bar Q_B)\}$. This in turn does not contain
any zero mode $(c_0-\bar c_0)$. Since both $\la \BB|$ and
$|\Lambda^{(0)}(p)\ra$ are annihilated by $(b_0-\bar b_0)$, we
see that the right hand side of \refb{ediff3} vanishes due to
the absence of the zero mode $(c_0-\bar c_0)$ in the matrix element.
This in turn establishes that the conserved charges $F^{(1)}$ and
$F^{(2)}$ are identical.

It also follows from the above analysis that if $|\Lambda\ra$ is
BRST trivial then the corresponding conserved charge vanishes. To
see this let us assume that $|\Lambda\ra = (Q_B+\bar Q_B)|\chi\ra$
for some ghost number zero state $|\chi\ra$. Let $|\chi(p)\ra$
denote a family of states labelled by the momentum $p$ such that
$|\chi(p=c)\ra=|\chi\ra$. Then the family
$|\wt\Lambda(p)\ra\equiv(Q_B+\bar Q_B)|\chi(p)\ra$ has the
property that it reduces to $|\Lambda\ra$ for $p=c$. Thus we can
compute the conserved charge associated with this symmetry using
this family $|\wt\Lambda(p)\ra$. However in this case $(Q_B+\bar
Q_B)|\wt\Lambda(p)\ra$ vanishes for all $p$, and hence the
corresponding state $|\wt\phi(p)\ra = (p-m)^{-1} (Q_B+\bar
Q_B)|\wt\Lambda(p)\ra$ also vanishes for all $p$. Using the
definition \refb{e9} of the conserved charge we see clearly that
the corresponding conserved charge also vanishes in this case.

Finally we note from the definition \refb{e9} of $F(x)$ and the
fact that $F(x)\propto e^{-icx}$ due to the conservation law, that
the value of $F$ depends on the matrix element $\la \BB|(c_0-\bar
c_0) |\phi(p)\ra$ at $p=c$. If $|\phi(p=c)\ra$ is BRST exact then
this matrix element vanishes and we do not get a non-trivial
conserved charge.

\sectiono{Symmetries and Conserved Charges in Two Dimensional String
Theory} \label{s3a}

In this section we shall use the results of section \ref{s2} to
construct infinite number of conserved charges in two dimensional
bosonic string theory. We begin with a brief review of two
dimensional string theory. The world-sheet description of the
theory involves a time like scalar field $X^0$, a Liouville field
theory with $c=25$ and the usual ghost fields $b$, $c$, $\bar b$,
$\bar c$. In the $\alpha'=1$ unit that we shall be using, the
Liouville theory is described by a single scalar field $\vp$ with
exponential potential in the world sheet action:\footnote{The
Liouvile theory with $c=25$ actually has a term $\propto \vp
e^{2\vp}$ in the world-sheet action\cite{polch,DDW}. As in
\cite{0304224,0305159,0305194} we shall regard the $c=25$
Liouville theory as the $c\to 25$ limit of theories with $c>25$.
For $c>25$, \refb{e15} (with $e^{2\vp}$ replaced by an appropriate
power of $e^\vp$) is the correct form of the action, but $\mu$
undergoes an infinite renormalization as we take the $c\to 25$
limit.}
 \be \label{e15}
s_{liouville}
= \int d^2 z \left({1\over 2\pi}\,
\p_z\vp \p_{\bar z}\vp + \mu\, e^{2\vp}\right)\, ,
 \ee
where $\mu$ is a
parameter which we shall set to unity by a shift of the field $\vp$.
There is also a linear dilaton field background along the Liouville
direction
 \be \label{edilaton}
\Phi_D = 2\vp\, ,
 \ee
which makes the total central charge associated with the Liouville
direction to be 25. For large negative $\vp$, the potential term
$e^{2\vp}$ is negligible, and $\vp$ behaves as a free scalar field
with background charge. This is also the region where the
effective string coupling constant $e^{\Phi_D}=e^{2\vp}$ is small.
We shall call this region the weak coupling region.

In order to understand the structure of BRST cohomology in the two
dimensional string theory, it will be convenient to first regard
the left and right moving components of the world-sheet scalar
field $X=iX^0$ and the Liouville field $\vp$ as independent,
construct states in the left- and the right-moving sectors
separately, and then combine them matching the momenta in the
left- and the right-moving sector to construct proper states of
the two dimensional string theory. We begin with the CFT
associated with the free scalar field $X$. Let us denote by $X_L$
and $X_R$ the left and the right-moving components of $X$. The
CFT, besides containing the usual primary states
$e^{ikX_L(0)}|0\ra_X$ and $e^{ikX_R(0)}|0\ra_X$, contains a set of
primaries $|j,m\ra_L$, $|j,m\ra_R$ of the form\cite{DVV}:
 \be \label{e12}
|j,m\ra_L = \PP^L_{j,m} e^{2im X_L(0)}|0\ra_X, \qquad |j,m\ra_R =
\PP^R_{j,m} e^{2im X_R(0)}|0\ra_X,
 \ee
where $\PP^L_{j,m}$ and $\PP^R_{j,m}$ are some combination of
non-zero mode $X_L$, $X_R$ oscillators of level $(j^2-m^2)$, and
$(j,m)$ are SU(2) quantum numbers with $-j\le m\le
j$.\footnote{The underlying SU(2) algebras in the left and
right-moving sectors of the world-sheet are generated by $e^{\pm
2iX_L}$, $i\p X_L$, $e^{\pm 2iX_R}$ and $i\bar\p X_R$. These
generators do not have well defined action on a vertex operator
$e^{ikX}$ for non-integer $k$, and hence do not generate a
symmetry of the theory unless $X$ is compactified on a circle of
self-dual radius. Nevertheless this algebra is useful in analyzing
the properties of the special primary states $|j,m\ra_X$ even when
$X$ represents a non-compact coordinate.} For example, we have
$|1,0\ra_L=\alpha_{-1}|0\ra_X$,
$|1,0\ra_R=\bar\alpha_{-1}|0\ra_X$, where $\alpha_n$, $\bar
\alpha_n$ are the usual oscillators of the $X$-field.
$\PP^L_{j,\pm j}$ and $\PP^R_{j,\pm j}$, being of level 0, must be
identity operators. Thus $|j,j\ra_L=e^{2ij X_L(0)}|0\ra$,
$|j,j\ra_R=e^{2ij X_R(0)}|0\ra$.

 We shall combine the left and the
right-moving modes to define:
 \be \label{exx2} |j,m\ra_X =
|j,m\ra_L \times |j,m\ra_R = \PP^L_{j,m} \PP^R_{j,m} \, e^{2i m
X(0)}|0\ra_X \, .
 \ee
 In fact this theory contains a more general set of primaries
 $|j,m\ra_L\times |j',m\ra_R$, but we shall not introduce a
 special symbol to label these states.
For later use we shall also define:
 \be \label{ejmap}
|j,m,p\ra_L = \PP^L_{j,m} \, e^{ipX_L(0)} |0\ra_X, \qquad
|j,m,p\ra_R = \PP^R_{j,m} \, e^{ipX_R(0)} |0\ra_X\, ,
 \ee
for an arbitrary $X$-momentum $p$, and
 \be \label{e13}
|j,m,p\ra_X = |j,m,p\ra_L \times |j,m,p\ra_R = \PP^L_{j,m}
\PP^R_{j,m} \, |p\ra_X\, ,
 \ee
 where
 \be \label{edefpx}
 |p\ra_X = e^{ipX(0)}|0\ra_X\, .
 \ee
We shall normalize $\PP^L_{j,m}$, $\PP^R_{j,m}$ such that:
 \be \label{e14}
~_X\la j,m,p|j',m, p'\ra_X = ~_X\la p|p'\ra_X \,  \delta_{jj'} =
2\pi\, \delta(p+p')\, \delta_{jj'},
 \ee
where $~_X\la\cdot|\cdot\ra_X$ denotes BPZ inner product in the
CFT of the $X$-field. The vanishing of this inner product for
$j\ne j'$ follows simply from the fact that the two states have
different conformal weights.

The Liouville field theory
contains a set of primary vertex operators $V_\beta$ of conformal
weight $(h_\beta, h_\beta)$ with
 \be \label{e16}
h_\beta = {1\over 4}
\beta (4-\beta)\, .
 \ee
For large negative $\vp$ where $\vp$ behaves as a free scalar
field with background charge, $V_\beta\sim e^{\beta\vp}$ for
$\beta< 2$. Although the world-sheet action is not that of a free
field theory, it describes a solvable
CFT\cite{goli,9206053,9403141,9506136,0104158}. There is a
reflection symmetry\cite{9403141,9506136} that relates the vertex
operator $V_\beta$ to $V_{4-\beta}$ up to a constant of
proportionality. Of these only the vertex operators $V_{2+iP}$ for
real $P$ describe $\delta$-function normalizable states. The
normalization of these vertex operators will be chosen such that
 \be \label{e16a}
\la V_{2 + iP}(1) V_{2+iP'}(0) \ra_{liouville}
= 2\pi \, \left[ \delta(P+P') - \left({\Gamma(iP)\over
\Gamma(-iP)}\right)^2 \delta(P-P')\right]\, .
 \ee
 The second term in this expression is required by the reflection
 symmetry\cite{9403141,9506136} $V_{2+iP} \equiv -\left({\Gamma(-iP)\over
\Gamma(iP)}\right)^2 \, V_{2-iP}$. With this normalization
 \be \label{evapprox}
 V_{2+iP} \, \, \sim \, \, e^{(2+iP)\vp} - \left({\Gamma(iP)\over
\Gamma(-iP)}\right)^2 \, e^{(2-iP)\vp}\, ,
 \ee
 for large negative $\vp$. The second term reflects the effect of
 the exponentially growing potential for large positive $\vp$.

 This normalization differs from the implicit normalization
 assumed in \cite{0402157} where the second delta function in
\refb{e16a} was absent, and we pretended that $V_{2+iP}$ and
$V_{2-iP}$ are independent vertex operators. As long as we work
with appropriate linear combinations of $V_{2+iP}$ and $V_{2-iP}$
which obey the reflection symmetry, this procedure gives the
correct result. The
 closest analogy of this in ordinary quantum mechanics is that
 while studying a free particle on a half line with
 Neumann (Dirichlet) boundary condition on the wave-function
 at the origin, we can
 study the theory on the full line with basis states $e^{ik x}$
 and at the end restrict the field configurations to be even (odd)
 under reflection around the origin. In contrast in this paper we
 use the convention that
$V_{2+iP}$ itself obeys the reflection symmetry dictated by the
CFT. This is analogous to using $2\, \cos x$ ($2\, \sin x$) as
basis functions for free particle on a half line with
 Neumann (Dirichlet) boundary condition on the wave-function
 at the origin.
 Any basis independent relation {\it e.g.} eq.\refb{eback1} will not
 be affected by this difference in the choice of basis.

For simplifying the notation we shall regard the vertex operator $V_\beta$
as a product of a left-chiral vertex operator $V^L_\beta$ and a
right-chiral vertex operator $V^R_\beta$ of dimension $(h_\beta,0)$ and
$(0,h_\beta)$ respectively, although in the final expression only the
product $V^L_\beta V^R_\beta$ will appear.

We shall now review some results on the chiral BRST cohomology of
the world-sheet
theory\cite{LIAN,MMS1,BMP,9108004,9201056,0307195,IMM,9209011} in
ghost number zero and one sectors.\footnote{Much of the earlier
\label{fo1} analysis on the BRST cohomology in two dimensional
string theory was done for $\mu=0$. We shall not try to carry out
a complete analysis of the BRST cohomology for $\mu\ne 0$. Instead
we shall use the elements of BRST cohomology found in earlier
analysis which do not require explicit use of the oscillators of
$\vp$, but involve only those states in the Liouville theory which
are obtained by the action of the Liouville Virasoro generators on
primary operators of the form $V_\alpha$. These states can be
easily defined in the interacting theory. Thus for example states
of the form $(c\p\vp(0) + \p c(0))|0\ra$, which are valid elements
of the BRST cohomology for $\mu=0$, will not be included in our
analysis since $\p\vp(0)|0\ra_{liouville}$ cannot be expressed as
a combination of Liouville Virasoro generators acting on a primary
state $V_\alpha(0)|0\ra_{liouville}$. On the other hand a state of
the form $c\p^2 c(0)|0\ra$, which is BRST invariant but was not
included as an element of the relative BRST cohomology since it is
proportional to $Q_B c\p\vp(0)|0\ra$, will now be regarded as a
valid element of the relative BRST cohomology. The other main
difference from the $\mu=0$ case is that for $\mu=0$ a state was
taken to vanish only if it vanishes when expressed in terms of the
Liouville oscillators. Since we are not making use of Liouville
oscillators, we shall consider a state to vanish if it is a linear
combination of null states of the Liouville Virasoro algebra.
Since the one point function of null states on a disk vanishes for
the D0-brane boundary CFT\cite{0101152}, this prescription does
not lead to any internal contradiction.}
 As we shall see, these will be
the basic building blocks for the construction of non-trivial
symmetry generators of the two dimensional string theory under
which D-branes are charged. For definiteness we shall describe the
results in the left-moving (holomorphic) sector, but identical
results hold in the right-moving sector as well. We begin in the
ghost number one sector. In this sector we have an infinite number
of elements of the BRST cohomology labelled by the SU(2) quantum
numbers $(j,m)$ with $-j\le m\le j$, represented by the states
 \ben \label{exx7}
| Y^L_{j,m}\ra &=&
|j,m\ra_L \otimes V^L_{2(1-j)}(0)|0\ra_{liouville} \otimes
c_1|0\ra_{ghost} \nonumber \\
 &=& \PP^L_{j,m}
e^{2imX_L(0)}|0\ra_X \otimes V^L_{2(1-j)}(0)|0\ra_{liouville}
\otimes c_1|0\ra_{ghost}\, ,
 \een
where $\PP^L_{j,m}$ has been defined in \refb{e12}. By
construction these states have zero $L_0$ eigenvalue. For later
use we define:
 \be \label{exx8}
| Y^L_{j,m}(p)\ra = \PP^L_{j,m} e^{ipX_L(0)}|0\ra_X \otimes
V^L_{2(1-j)}(0)|0\ra_{liouville} \otimes c_1|0\ra_{ghost}\, .
 \ee

In the ghost number 0 sector also we have an infinite number of
elements of the BRST cohomology labelled by the SU(2) quantum
numbers $(j-1,m)$ with $-(j-1)\le m\le j-1$. The representative
elements of the BRST cohomology can be chosen to be of the form
 \be \label{exx5-}
|\OO^L_{j-1,m}\ra = \QQ^L_{j-1,m}
|j-1,m\ra_L \otimes V^L_{2(1-j)}(0)|0\ra_{liouville} \otimes
c_1|0\ra_{ghost}\, ,
 \ee
where $\QQ^L_{j-1,m}$ is an operator of ghost number $-1$, level
$(2j-1)$ constructed from negative moded ghost oscillators and $X$
and Liouville Virasoro generators\cite{IMM}. Using \refb{e12} this
can be rewritten as
 \be \label{exx5--}
|\OO^L_{j-1,m}\ra = \RR^L_{j-1,m}
e^{2imX_L(0)}|0\ra_X \otimes V^L_{2(1-j)}(0)|0\ra_{liouville} \otimes
c_1|0\ra_{ghost}
 \ee
where $\RR^L_{j-1,m}\equiv \QQ^L_{j-1,m}\, \PP^L_{j-1,m}$ is an
operator of ghost number $-1$ constructed from negative moded $X$
and ghost oscillators and Liouville Virasoro
generators.\footnote{Note that while $\QQ^L_{j-1,m}$ is built from
$X$ Virasoro generators, $\RR^L_{j-1, m}$ is built from $X$
oscillators. The distinction is important. While any combination
of $X$ Virasoro generators, acting on a state of definite
$X$-momentum, can be expressed in terms of $X$-oscillators, the
reverse is not always true.}
 For later use we now define:
 \be \label{exx5}
|\OO^L_{j-1,m}(p)\ra = \RR^L_{j-1,m} e^{i p X_L(0)}|0\ra_X
\otimes V^L_{2(1-j)}(0)|0\ra_{liouville} \otimes
c_1|0\ra_{ghost}
\, .
 \ee
Note that $|Y^L_{j,m}(p)\ra$ and $|\OO^L_{j-1,m}(p)\ra$ are both
built by the action of $X$ and ghost oscillators and Liouville
Virasoro generators on the same Fock vacuum $e^{ipX_L(0)}|0\ra_X
\otimes V^L_{2(1-j)}(0)|0\ra_{liouville} \otimes
c_1|0\ra_{ghost}$, and satisfy
 \be \label{eb0}
b_0 |\OO^L_{j-1,m}(p)\ra = 0, \qquad b_0 | Y^L_{j,m}(p)\ra = 0\, .
 \ee
Furthermore, since $|Y^L_{j,m}(p=2m)\ra$ and
$|\OO^L_{j-1,m}(p=2m)\ra$ have zero $L_0$ eigenvalues, we have
 \be \label{eloev}
L_0|\OO^L_{j-1,m}(p)\ra = {1\over 4}\, (p^2 - 4m^2) \,
|\OO^L_{j-1,m}(p)\ra\, , \qquad L_0 |Y^L_{j,m}(p)\ra = {1\over 4}
\, (p^2 - 4m^2) \, |Y^L_{j,m}(p)\ra\, . \ee

Given that $|\OO^L_{j-1,m}\ra=|\OO^L_{j-1,m}(p=2m)\ra$ and
$| Y^L_{j,m}\ra=| Y^L_{j,m}(p=2m)\ra$ are
BRST invariant, we must
have\footnote{For a chiral state
we shall use the convention that at $p=2m$ a
state labelled by
the subscript $_{j,m}$ carries definite SU(2) quantum numbers $(j,m)$,
whereas a state labelled be the subscript $_{(j),m}$ does not in general
have definite SU(2) quantum numbers. After combining the left- and the
right-moving sectors to get full closed string states we shall no longer
follow this convention.}
 \be \label{exx6}
Q_B |\OO^L_{j-1,m}(p)\ra = (p-2m) |\eta^L_{(j),m}(p)\ra \, ,
 \ee
and
 \be \label{exx9}
Q_B | Y^L_{j,m}(p)\ra = (p-2m) |\psi^L_{(j),m}(p)\ra\, ,
 \ee
for some states $|\eta^L_{(j),m}(p)\ra$ and $|\psi^L_{(j),m}(p)\ra$.
It follows from eqs.\refb{exx6} and \refb{exx9} and the nilpotence of
$Q_B$ that both $|\eta^L_{(j),m}(p)\ra$ and $|\psi^L_{(j),m}(p)\ra$ are
BRST invariant for any $p\ne 2m$ and hence by analytic continuation
also for $p=2m$. We also see from eqs.\refb{exx6}, \refb{exx9} that
for $p\ne 2m$, $|\eta^L_{(j),m}(p)\ra$ and $|\psi^L_{(j),m}(p)\ra$ are
BRST trivial but for $p=2m$ they can be BRST non-trivial. Finally we
note that
since $|\OO^L_{j-1,m}(p)\ra$ and $| Y^L_{j,m}(p)\ra$ have
non-vanishing $L_0$ eigenvalues proportional to $(p^2-4m^2)$,
$|\eta^L_{(j),m}(p)\ra$
and $|\psi^L_{(j),m}(p)\ra$
defined through \refb{exx6} and \refb{exx9}
are not annihilated by $b_0$ in general.

It has been shown in appendix \ref{sappa} that at $p=2m$,
 \be \label{exi1}
|\eta^L_{(j),m}\ra \equiv |\eta^L_{(j),m}(p=2m)\ra  = | Y^L_{j,m}\ra
+ |\wh \eta^L_{(j),m}\ra \, ,
 \ee
 \be \label{esp0}
|\psi^L_{(j),m}\ra \equiv |\psi^L_{(j),m}(p=2m)\ra  = m \, c_0\,
| Y^L_{j,m}\ra
+ |\tau^L_{j-1,m}\ra\, ,
 \ee
where $|\wh \eta^L_{(j),m}\ra$ is a linear combination of states
carrying SU(2) quantum numbers $(j-1,m)$ and $(j-2,m)$, and
$|\tau^L_{j-1,m}\ra$ has SU(2) quantum numbers $(j-1,m)$.

This finishes our discussion of chiral BRST cohomology in ghost
numbers zero and one sectors. We shall now combine the left and
the right-moving states matching $X$ and $\vp$ momenta to
construct a family of states $|\Lambda_{j,m}(p)\ra$ of ghost
number 1, satisfying the requirement \refb{e5}. We
define:\footnote{Note that these gauge transformation parameters
are of order $e^{2(1-j)\vp}$ for large negative $\vp$ and hence
grow exponentially in this region for $j>1$. Nevertheless, since
the D0-brane is located in the strong coupling region of $\vp> 0$,
we expect these transformations to describe sensible symmetries of
the open string field theory on the D0-brane.}
 \ben \label{exx11}
|\Lambda_{j,m}(p)\ra = {1\over 2} \left[ |\OO^L_{j-1,m}(p)\ra
\times | Y^R_{j,m}(p)\ra - | Y^L_{j,m}(p)\ra\times
|\OO^R_{j-1,m}(p)\ra
\right], \nonumber \\
\quad j\ge 1, \quad -(j-1)\le m\le j-1 \, . \quad ~
 \een
{}From \refb{eb0}, \refb{eloev}, their right-moving counterpart,
it follows that
 \be \label{exx14}
(b_0-\bar b_0) |\Lambda_{j,m}(p)\ra = (L_0-\bar L_0)  |\Lambda_{j,m}(p)\ra
= 0\, ,
 \ee
Also using \refb{exx6}, \refb{exx9} and their right-moving counterpart,
we get:
 \be \label{exx12}
(Q_B+\bar Q_B) |\Lambda_{j,m}(p)\ra = (p-2m) |\phi_{j,m}(p)\ra \, ,
 \ee
where
 \ben \label{exx13}
|\phi_{j,m}(p) \ra &=& {1\over 2}
\bigg[ |\eta^L_{(j),m}(p)\ra\times
| Y^R_{j,m}(p)\ra
+|\OO^L_{j-1,m}(p)\ra \times
|\psi^R_{(j),m}(p)\ra \nonumber \\
 && - |\psi^L_{(j),m}(p)\ra \times
|\OO^R_{j-1,m}(p)\ra  + | Y^L_{j,m}(p)\ra \times
|\eta^R_{(j),m}(p)\ra
\bigg]\, .
 \een
Eqs.\refb{exx14}, \refb{exx12} now give
 \be \label{exx14a}
(b_0-\bar b_0) |\phi_{j,m}(p)\ra = 0, \qquad (L_0-\bar L_0)
|\phi_{j,m}(p)\ra = 0
\, .
 \ee

For explicit computation of the conserved charge in section
\ref{s3b} we shall need the form of $|\phi_{j,m}(p=2m)\ra$. Using
eqs.\refb{exi1}, \refb{esp0} we get
 \ben \label{ephijmp}
|\phi_{j,m}(p=2m)\ra &=& | Y^L_{j,m}\ra \times | Y^R_{j,m}\ra +
|\omega_{j,m}\ra \nonumber \\
&& +{1\over 2} \left[  |\wh \eta^L_{(j),m}\ra \times  |
Y^R_{j,m}\ra +  |
Y^L_{j,m}\ra \times  |\wh \eta^R_{(j),m}\ra\right] \nonumber \\
&& +{m\over 2} \, \left[ |\OO^L_{j-1,m}\ra \times \bar c_0  |
Y^R_{j,m}\ra - c_0 | Y^L_{j,m}\ra \times   |\OO^R_{j-1,m}\ra
\right]\, ,\een
 where
 \ben \label{exx25b}
|\omega_{j,m}\ra &=& {1\over 2}  |\OO^L_{j-1,m}\ra \times
|\tau^R_{j-1,m}\ra - {1\over 2}  |\tau^L_{j-1,m}\ra \times
|\OO^R_{j-1,m}\ra \nonumber \\
&=& {1\over 2\sqrt 2} \, \left( |\OO^L_{j-1,m}\ra \times
\sum_{n=1}^\infty \bar c_{-n} \bar \alpha_n |Y^R_{j,m}\ra -
\sum_{n=1}^\infty c_{-n} \alpha_n |Y^L_{j,m}\ra\times
|\OO^R_{j-1,m}\ra\right) \, .
 \een
In going from the first to the second line of \refb{exx25b} we
 have used eq.\refb{espp1}.
The first term in \refb{exx25a} is built on the primary
$|j,m\ra_{X}$, whereas the second term $|\omega_{j,m}\ra$ is built
on the primary $|j-1,m\ra_{X}$. $|\omega_{j,m}\ra$ are BRST
invariant ghost number two closed string states, carrying
$SU(2)_L\times SU(2)_R$ quantum numbers $(j-1,m;j-1,m)$, which are
not of the form $|Y^L_{j-1,m}\ra \times |Y^R_{j-1,m}\ra$. They
represent elements of the relative BRST cohomology which cannot be
factored into a ghost number 1 state on the left and a ghost
number 1 state on the right. For example $|\omega_{1,0}\ra\propto
(c_{-1} c_1 - \bar c_{-1}\bar c_1)|0\ra$.\footnote{As pointed out
in footnote \ref{fo1}, this state is BRST trivial for $\mu=0$, but
not for $\mu\ne 0$.  Physically this reflects that a constant
dilaton, which was a pure gauge deformation for $\mu=0$ since it
could be absorbed by translating $\vp$, is no longer a pure gauge
for $\mu\ne 0$ since translation of $\vp$ now will also change
$\mu$. In fact one can explicitly check that the vertex operator
associated with the state $(c_0-\bar c_0)|\omega_{1,0}\ra$ has a
non-vanishing inner product with the boundary state describing a
D0-brane.} In general $|\omega_{j,m}\ra$ will be proportional to
the BRST invariant state $\sum_n n(c_{-n}c_n - \bar c_{-n}\bar
c_n)|\OO^L_{j-1,m}\ra \times |\OO^R_{j-1,m}\ra$\cite{IMM} up to
addition of BRST trivial terms.

Eqs.\refb{e9}-\refb{e11} now tell us that for a D-brane in this
two dimensional string theory described by a boundary state
$|\BB\ra$,
 \be \label{e19}
\p_x \left(e^{2im x} F_{j,m} (x) \right) =
0\, ,
 \ee where
 \be \label{e20}
F_{j,m}(x) = \int {dp\over 2\pi} \, e^{-i
p x} \la \BB | (c_0-\bar c_0)|\phi_{j,m}(p)  \ra\, .
 \ee
Since \refb{e19} shows that $F_{j,m}(x)\propto e^{-2imx}$, the
contribution to \refb{e20} must come from a term proportional to
$\delta(p-2m)$ in the integrand. Thus computation of the actual
value of $F_{j,m}(x)$ should only involve $|\phi_{j,m}(p)\ra$
evaluated at $p=2m$, despite the fact that its definition in terms
of the boundary state involves $|\phi_{j,m}(p)\ra$ for general
$p$.

The Minkowski
continuation $x\to i x^0$ then gives $e^{-2mx^0}F_{j,m}(ix^0)$ as the
conserved charge. We define
 \be \label{edefqjm}
Q_{j,m}(x^0) = -{1\over 2\sqrt \pi} F_{j,m}(ix^0) \, ,
 \ee
where the overall normalization constant $-{1\over 2\sqrt \pi}$
has been fixed arbitrarily in order to simplify some of the
formul\ae\ which will appear later. $e^{-2mx^0} Q_{j,m}(x^0)$
gives the conserved charge in the Minkowski theory associated with
the global symmetry generated by $|\Lambda_{j,m}\ra$. We shall
loosely refer to $Q_{j,m}$ as the conserved charge with the
understanding that it is $e^{-2mx^0} Q_{j,m}(x^0)$ that is really
conserved. We could have absorbed the factor $e^{-2mx^0}$ in the
definition of $Q_{j,m}$ so that $Q_{j,m}$ itself would be
conserved. However the definition \refb{edefqjm} has the advantage
that it does not involve any explicit time dependent factors. This
guarantees that when we relate $Q_{j,m}$ to some quantity in the
matrix model, then it must be formed out of matrix model variables
without any explicit time dependent factors.

Before concluding this section we shall illustrate our result with
a simple example. For $(j,m)=(1,0)$ the symmetry generator
$|\Lambda_{1,0}\ra$ takes the form
 \be \label{esy1}
|\Lambda_{1,0}\ra = (c_1 \alpha_{-1} - \bar c_1 \bar
\alpha_{-1})|0\ra
 \ee
 up to an overall normalization constant.
This represents translation along $x$. We define
 \be \label{esy2}
|\Lambda_{1,0}(p)\ra = (c_1 \alpha_{-1} - \bar c_1 \bar
\alpha_{-1})|p\ra\, ,
 \ee
where $|p\ra = e^{ip X(0)}|0\ra$. Using the definition
\refb{exx12} and the form of $Q_B$ given in \refb{esp2} we get
 \be \label{esy3}
|\phi_{1,0}(p)\ra = {1\over \sqrt 2} \left[ {p\over 2\sqrt 2}
(c_0+\bar c_0) \, (c_1 \alpha_{-1} - \bar c_1 \bar \alpha_{-1}) -2
c_1 \bar c_1 \alpha_{-1} \bar \alpha_{-1} + (c_{-1} c_1 - \bar
c_{-1} \bar c_1)\right] |p\ra\, . \ee We can use this to define
$F_{1,0}(x)$. However as argued earlier, the actual value of
$F_{1,0}$ can only depend on $|\phi_{1,0}(p=0)\ra$, and comes from
a term in $|\BB\ra$ built over the $|p=0\ra$ state:
 \be \label{esy4}
 (c_0+\bar c_0) \left[ A c_1 \bar c_1 \alpha_{-1} \bar
\alpha_{-1} + B (c_{-1} c_1 - \bar c_{-1} \bar c_1 ) + \ldots
\right] |0\ra \, ,
 \ee
  where $\ldots$ involves higher level
oscillators. \refb{e20}, \refb{esy3} now gives:
 \be \label{esy5} F_{1,0}
\propto (A + B) \, .
 \ee
 Upon continuation to Minkowski space this
agrees precisely with the expression for the energy of a D-brane
as given in \cite{0203265}.

\sectiono{Charges Carried by the Rolling
Tachyon Background
} \label{s3b}

Two dimensional string theory has a D0-brane, described by the
boundary state
 \be \label{ebzero}
 |\BB_0\ra = |\BB_0\ra_X \otimes |\BB\ra_{liouville} \otimes
|\BB\ra_{ghost}\, .
 \ee
Here $|\BB\ra_{ghost}$ denotes the standard boundary state in the
ghost sector that accompanies any D-brane
 \be \label{ebghost}
|\BB\ra_{ghost} = \exp\left( -\sum_{n=1}^\infty (\bar b_{-n}
c_{-n} + b_{-n} \bar c_{-n})\right) (c_0+\bar c_0) c_1\bar c_1
|0\ra_{ghost}\, .
 \ee
$|\BB\ra_{liouville}$ denotes the boundary state of the D0-brane
CFT in the Liouville sector\cite{0101152,0305159}:
 \be \label{ebliou}
|\BB\ra_{liouville} = -{i\over g_s} \, {1\over 2\sqrt\pi}
\int{dP\over 2\pi} \, \sinh(\pi P) \, {\Gamma(-iP)\over
\Gamma(iP)} \, |P\ra\ra_{liouville}\, ,
 \ee
where $|P\ra\ra_{liouville}$ is the Virasoro Ishibashi
state\cite{ishibashi} built on the primary
$|P\ra_{liouville}\equiv
V_{2+iP}(0)|0\ra_{liouville}$.\footnote{Note that
$|P=0\ra_{liouville}\ne |0\ra_{liouville}$. We apologize for this
inconsistency in notation.} Finally $|\BB_0\ra_X$ is given by:
 \be \label{ebox}
 |\BB_0\ra_X=\exp\left(-\sum_{n=1}^\infty {1\over n} \alpha_{-n}\bar
 \alpha_{-n} \right) |0\ra_X\, .
 \ee
 Besides an overall factor of $g_s^{-1}$, the boundary state
 $|\BB_0\ra$ apparently differs from the corresponding boundary
 state given in \cite{0402157} by a factor of ${1\over 2}$. This
 is a reflection of the difference in the choice of convention for
 $V_{2+iP}$ as mentioned below eq.\refb{evapprox}. Alternatively
 we could multiply the right hand side of \refb{ebliou} by a factor of
 2, and restrict the integration over $P$ to positive real axis
 only.

The open string spectrum on this D0-brane has a tachyon. We can
construct a family of classical solutions describing the rolling
of this tachyon away from the maximum of the tachyon
potential\cite{0203211,0203265}. In the euclidean version the
family of boundary CFT's describing these solutions is obtained by
deforming the original CFT of the D0-brane by the boundary term
 \be
\label{e20a} \tl\, \int \, dt\, \cos(X(t))\, ,
 \ee
where $t$ denotes a parameter labelling the boundary of the
world-sheet and $\tl$ is the parameter labelling the rolling
tachyon solution. The boundary state $|\BB\ra$ associated with
this classical solution has the form:
 \be \label{eb1}
|\BB\ra = |\BB_1\ra + |\BB_2\ra
 \ee
where
 \ben \label{eb2}
|\BB_1\ra &=& |\BB_1\ra_{X} \otimes
|\BB\ra_{liouville} \otimes |\BB\ra_{ghost}
\nonumber \\
|\BB_2\ra &=& |\BB_2\ra_{X} \otimes
|\BB\ra_{liouville} \otimes |\BB\ra_{ghost}
\, .
 \een
$|\BB\ra_{ghost}$ and $|\BB\ra_{liouville}$ are as given in
\refb{ebghost}, \refb{ebliou}. Computation of
$|\BB_1\ra_X+|\BB_2\ra_X$ has been given in
refs.\cite{9402113,9811237,0208196,0212248,0305177,0402157,0301038,0308172}.
Here we shall follow the approach of ref.\cite{0402157} and
express the total boundary state as a sum of two parts.
$|\BB_1\ra_{X}$ is given by
 \be \label{eb3}
|\BB_1\ra_{X} = \exp\left(\sum_{n=1}^\infty {1\over n} \alpha_{-n}
\bar\alpha_{-n}\right)\,f(X(0))|0\ra_{X}\,  ,
 \ee
where
 \be \label{eb3a}
\qquad f(x) = {1\over 1 + \sin(\pi\tl)e^{ix}} + {1\over 1 +
\sin(\pi\tl)e^{-ix}}-1= \sum_{2m\in Z} (-1)^{2m} \,
\sin^{2|m|}(\pi\tl) \, e^{2imx}\, .
 \ee
On the other hand $|\BB_2\ra_{X}$ is given by:
 \be \label{eb4}
|\BB_2\ra_{X} = \sum_{j\ge 1}
\sum_{m=1-j}^{j-1} f_{j,m}(\tl) |j,m\ra\ra_{X}
 \ee
where $|j,m\ra\ra_{X}$ denotes the Virasoro Ishibashi
state\cite{ishibashi} built over the primary state $|j,m\ra_{X}$,
and
 \be \label{e22}
f_{j,m}(\tl) = D^j_{m,-m}(2\pi \tl) {(-1)^{2m}\over
D^j_{m,-m}(\pi)} - (-1)^{2m} \sin^{2|m|}(\pi\tl)\, .
 \ee
Here $D^j_{m,n}(\theta)$ denotes the representation of the SU(2)
group element $e^{i\theta\sigma_1/2}$ in the spin $j$
representation. Since $f_{j,m}(\tl)$ involves only the ratio
$D^j_{m,-m}(2\pi \tl)/D^j_{m,-m}(\pi)$, it is independent of the
choice of the phase of the basis states used to define
$D^j_{m,n}$. With some particular choice of the phases of the
basis states we have (see {\it e.g.} \cite{9811237})
 \ben
\label{edjm}
D^j_{m,n}(2\pi\tl) &=& \sum_{\mu = max(0,n-m)}^{min(j-m,j+n)}
{[(j+m)!(j-m)!(j+n)!(j-n)!]^{1\over 2} \over (j-m-\mu)!(j+n-\mu)! \mu!
(m-n+\mu)!} \nonumber \\
&& \, (\cos(\pi\tl))^{2(j-\mu)+n-m}
(i\sin(\pi\tl))^{m-n+2\mu}\, .
 \een
Using \refb{edjm} and some algebra,
eq.\refb{e22} may be rewritten as
 \be \label{efjm}
f_{j,m}(\tl) = (-1)^{2|m|} \cos^2(\pi\tl) \sin^{2|m|}(\pi\tl)
\sum_{s=0}^{j-|m|-1} \, \alpha^s_{j,m} \, \sin^{2s}(\pi\tl)\, ,
 \ee
where
 \ben \label{efjm1}
\alpha^s_{j,m} &\equiv& \sum_{\mu=0}^s {(j+|m|)! (j-|m|)!
(j-|m|-1-\mu)! (-1)^{s-j+|m|} \over ((j-|m|-\mu)!)^2 \mu!
(2|m|+\mu)! (s-\mu)! (j-|m|-1-s)!} -
1 \nonumber \\
&& \qquad \qquad \qquad \qquad \qquad \qquad
\qquad \qquad \qquad \qquad \qquad \qquad \hbox{for $s\le j-|m|-1$}\, ,
\nonumber \\
&\equiv& 0 \quad \hbox{for $s > j-|m|-1$} \, .
 \een
Thus $f_{j,m}(\tl)$ depends on the magnitude of $m$ but not its
sign.

$|\BB_1\ra_{X}$ given in \refb{eb3} can also be expressed in terms
of the Ishibashi states $|j,m\ra\ra_{X}$ as follows. We rewrite
\refb{eb3} as
 \ben \label{erew1}
|\BB_1\ra_{X} &=& \sum_{2m\in Z}
\exp\left(\sum_{n=1}^\infty{1\over n}
\alpha_{-n}\bar\alpha_{-n}\right) \, (-1)^{2m} \sin^{2|m|}(\pi\tl)
e^{2imX(0)}|0\ra \nonumber \\
&=& \sum_{2m\in Z} \, (-1)^{2m} \sin^{2|m|}(\pi\tl) \, \sum_N
|N,m\ra_L \otimes |N,m\ra_R \, ,
 \een
where $|N,m\ra_L$ ($|N,m\ra_R$) denote a complete basis of
orthonormal states constructed out of products of $\alpha_{-n}$'s
($\bar\alpha_{-n}$'s) acting on $e^{2imX_L(0)}|0\ra$
($e^{2imX_R(0)}|0\ra$). In going from the first to the second line
of \refb{erew1} we have expanded
$\exp\left(\sum_{n=1}^\infty{1\over n}
\alpha_{-n}\bar\alpha_{-n}\right)$ in a power series expansion. If
$\wt{|N,m\ra_L}$ ($\wt{|N,m\ra_R}$) denote another basis of
orthonormal states, related to the previous basis by a rotation:
 \be \label{erew2}
\wt{|N,m\ra_L} = R_{NN'} |N',m\ra_L, \quad
\wt{|N,m\ra_R} = R_{NN'} |N',m\ra_R\, ,
\quad R_{NM}R_{NM'}=\delta_{MM'}\, ,
 \ee
then \refb{erew1} may be reexpressed as
 \be \label{erew3}
|\BB_1\ra_{X} = \sum_{2m\in Z} \, (-1)^{2m} \sin^{2|m|}(\pi\tl) \,
\sum_N \wt{|N,m\ra_L} \otimes \wt{|N,m\ra_R} \, .
 \ee
Choosing $\wt{|N,m\ra_L}$ and $\wt{|N,m\ra_R}$ to be the
orthonormal basis of states formed out of $|j,m\ra_L$, $|j,m\ra_R$
and their Virasoro descendants, we arrive at the equation:
 \be \label{erew4}
|\BB_1\ra_{X} = \sum_{j\ge 0} \, \sum_{m=-j}^j \, (-1)^{2m}
\sin^{2|m|}(\pi\tl) \, |j,m\ra\ra_X\, .
 \ee

We now turn to the computation of \refb{e20} for the boundary state
$|\BB\ra$ described above. The first important point to note is that the
contribution to \refb{e20} from the boundary state $|\BB_1\ra$ vanishes.
To see this we note that the state:
 \be \label{exx21}
|\BB_1(p)\ra \equiv \exp\left(\sum_{n=1}^\infty {1\over n} \alpha_{-n}
\bar\alpha_{-n}\right) e^{ip.X(0)}|0\ra_{X}
\otimes |\BB\ra_{liouville}\otimes |\BB\ra_{ghost}
 \ee
is BRST invariant for every $p$\cite{0402157}:
 \be \label{exx22}
(Q_B+\bar Q_B) |\BB_1(p)\ra = 0\, .
 \ee
As a result, using \refb{exx12} we get
 \be \label{exx23a}
\la \BB_1(p')| (c_0-\bar c_0) |\phi_{j,m}(p)\ra =
(p-2m)^{-1}\, \la \BB_1(p')| \{(c_0-\bar
c_0), (Q_B+\bar Q_B)\} |\Lambda_{j,m}(p)\ra \, .
 \ee
Now $\{(c_0-\bar
c_0), (Q_B+\bar Q_B)\}$ does not have any $c$ or $\bar c$ zero mode. Using
\refb{exx14} we also see that $|\Lambda_{j,m}(p)\ra$ does not have any
$(c_0-\bar c_0)$ zero mode. On the other hand the $|\BB\ra_{ghost}$
appearing in the expression for $|\BB_1\ra$ contains only a zero mode of
$(c_0+\bar c_0)$ but not of $(c_0-\bar c_0)$. As a result the operators
appearing in the matrix
element \refb{exx23a} does not have any $(c_0-\bar c_0)$ zero mode and
hence the matrix element vanishes. This gives:
 \be \label{exx23}
\la \BB_1(p')| (c_0-\bar c_0) |\phi_{j,m}(p)\ra = 0\, ,
 \ee
for every $p\ne 2m$. By analytic continuation\footnote{The inner
product in the left hand side of \refb{exx23} carries a factor of
$2\pi \delta(p+p')$ which is clearly not an analytic function of
$p$. Eq.\refb{exx23} implies that the coefficient of this
$\delta$-function, regarded as a function of $p$ after setting
$p'=-p$, vanishes. It is this coefficient that is expected to be
an analytic function of $p$.} this result must hold also for
$p=2m$. Since $|\BB_1\ra$ defined in \refb{eb2}, \refb{eb3} is a
linear combination of $|\BB_1(p)\ra$ defined in \refb{exx21}, we
see that
 \be \label{exx24}
\la \BB_1| (c_0-\bar c_0) |\phi_{j,m}(p)\ra = 0  \, .
 \ee
This agrees with the conclusion of \cite{0402157} that the $|\BB_1\ra$
part of the boundary state does not carry any information about conserved
charges.

Thus we have
 \be \label{exx25}
F_{j,m}(x) = \int {dp\over 2\pi} \, e^{-i
p x} \la \BB_2 | (c_0-\bar c_0)|\phi_{j,m}(p)  \ra\, .
 \ee
As pointed out below \refb{e20}, the contribution to $F_{j,m}(x)$
only involves $|\phi_{j,m}(p=2m)\ra$. Furthermore, since
$|\BB_2\ra_{X}$ contains contribution from states built on
primaries of the form $|j,m\ra_L\times |j,m\ra_R$, we can throw
away from $|\phi_{j,m}(p=2m)\ra$ all terms which are built on
primaries of the form $|j',m\ra_L\times |j'',m\ra_R$ with $j'\ne
j''$. The part of $|\phi_{j,m}(p=2m)\ra$ that contributes to
\refb{exx25} is given by the terms in the first line on the right
hand side of \refb{ephijmp}:
 \be \label{exx25a}
|Y^L_{j,m}\ra\times |Y^R_{j,m}\ra + |\omega_{j,m}\ra\, ,
 \ee
 since $|\wh\eta^L_{(j),m}\ra$ ($|\wh\eta^R_{(j),m}\ra$) is a
 linear combination of states with left (right) SU(2) quantum
 numbers $(j-1,m)$ and $(j-2,m)$.
The contribution to \refb{exx25} from the first term of
\refb{exx25a} is easy to evaluate. The relevant part of the
boundary state $|\BB_2\ra$ that contributes to this matrix element
is
 \be \label{e21}
f_{j,-m}(\tl) |j,-m\ra_{X} \otimes |\BB\ra_{liouville} \otimes
(c_0+\bar c_0)c_1\bar c_1|0\ra_{ghost}
 \ee
Using eq.\refb{e14}, \refb{exx8} and the normalization condition
 \be \label{e23}
\la 0| c_{-1}\bar c_{-1} c_0 \bar c_0 c_1\bar
c_1|0\ra_{ghost}=1\, ,
 \ee
we get the contribution to \refb{exx25} from this term to be
 \be \label{e24a}
2 \, f_{j,-m}(\tl) \, e^{-2imx} \, ~_{liouville}\la \BB|
V_{2(1-j)}(0) | 0\ra_{liouville} \, .
 \ee

Since $|\omega_{j,m}\ra$ carries left and right SU(2) quantum
numbers $(j-1,m)$, the contribution to \refb{exx25} from the
second term on the right hand side of \refb{exx25a} will be
proportional to $2f_{j-1,-m}(\tl) \, e^{-2imx} \, ~_{liouville}\la
\BB| V_{2(1-j)}(0) | 0\ra_{liouville}$. If we denote the constant
of proportionality by $a_{j,m}$, then this contribution is
 \be \label{e24b}
2 \, a_{j,m} \, f_{j-1,-m}(\tl) \, e^{-2imx} \, ~_{liouville}\la
\BB| V_{2(1-j)}(0) | 0\ra_{liouville}\, .
 \ee
Combining \refb{e24a} and \refb{e24b}
we now get
 \be \label{e24-}
F_{j,m}(x) = 2 \, (f_{j,-m}(\tl) + a_{j,m} \, f_{j-1,-m}(\tl)) \,
e^{-2imx} \, ~_{liouville}\la \BB| V_{2(1-j)}(0) |
0\ra_{liouville} \, .
 \ee

While a direct computation of $a_{j,m}$ is somewhat involved we can
calculate it indirectly by using \refb{exx24} and
the form of
$|\BB_1\ra_X$ given in \refb{erew4}.
Using the same line of argument as used for $|\BB_2\ra$,
we get the contribution to
$F_{j,m}(x)$ from $|\BB_1\ra$ to be
 \be \label{ebb11}
2 \, (-1)^{2m} \, \sin^{2|m|}(\pi\tl) \, (1+a_{j,m}) \, \,
e^{-2imx} \, ~_{liouville}\la \BB| V_{2(1-j)}(0) |
0\ra_{liouville} \, .
 \ee
On the other hand from the general argument given earlier we know
that the contribution to $F_{j,m}(x)$ from $|\BB_1\ra$ must vanish.
This gives:
 \be \label{eajm}
a_{j,m}=-1\, .
 \ee
Hence \refb{e24-} takes the form:
 \be \label{e24}
F_{j,m}(x) = 2 \, (f_{j,-m}(\tl) - f_{j-1,-m}(\tl)) \, e^{-2imx}
\, ~_{liouville}\la \BB| V_{2(1-j)}(0) | 0\ra_{liouville} \, .
 \ee

Using eqs.\refb{e16a}, \refb{ebliou} we get
 \be \label{e26}
 ~_{liouville}\la \BB|
V_{2+iP}(0) | 0\ra_{liouville}  = {1\over g_s} \, {1\over
\sqrt\pi} \, i \sinh(\pi P) \, {\Gamma(iP)\over \Gamma(-iP)}\, .
 \ee
Taking $P\to 2ij$ limit in the above formula we get
 \be \label{e28}
 ~_{liouville}\la \BB|
V_{2(1-j)}(0) | 0\ra_{liouville} = {1\over g_s} \, {\sqrt \pi\over
(2j)!(2j-1)!}\, .
 \ee
Thus
 \be \label{e29}
F_{j,m}(x) = {2 \,\sqrt \pi\over (2j)!(2j-1)!}\,{1\over g_s} \,
(f_{j,-m}(\tl) - f_{j-1,-m}(\tl)) \, e^{-2imx} \, .
 \ee
Under
the replacement $x\to ix^0$ this gives
 \be \label{e30}
Q_{j,m}(x^0)\equiv -{1\over 2\sqrt\pi} \, F_{j,m}(i x^0) =-
{1\over g_s}\, {1 \over (2j)!(2j-1)!}\, (f_{j,-m}(\tl) -
f_{j-1,-m}(\tl)) \, e^{2mx^0} \, ,
 \ee
 Using \refb{efjm} we can rewrite this as
 \be \label{enewqjm}
Q_{j,m}(x^0) = {1\over g_s}\, (-1)^{2m} \, e^{2m x^0} \,
\cos^2(\pi\tl)  \sum_{l=|m|}^{j-1} b^{l,m}_{j} \sin^{2l}(\pi\tl)\,
,
 \ee
where
 \ben \label{e38a}
b^{l,m}_{j} &=& -{1\over (2j)!(2j-1)!}\, \left[
\alpha^{l-|m|}_{j,-m} - \alpha^{l-|m|}_{j-1,-m} \right]
\quad \hbox{for $|m|\le l\le j-1$}\, , \nonumber \\
&=& 0 \quad \hbox{otherwise}\, .
 \een
The coefficients $\alpha^s_{j,m}$ have been defined in \refb{efjm1}.
It is understood that the sum over $l$ in \refb{enewqjm} will always
be in integer steps but $l$ itself can be integer or half integer
depending on whether $m$ is integer or half integer.

 {}From \refb{enewqjm}, \refb{e38a}, \refb{efjm1} we get
 \be \label{eextraqjm}
  Q_{1,0} = {1\over g_s}\cos^2(\pi\tl) \,
b^{0,0}_1 = {1\over g_s} \, \cos^2(\pi\tl)\, .
 \ee
 Thus $Q_{1,0}$ measures the total energy of the
 D0-brane\cite{0203211,0203265}.

Although the above analysis defines infinite number of conserved
charges, for a single D0-brane information about these conserved
charges is highly superfluous. Since the classical trajectories of
a D0-brane are labelled by a single parameter $\tl$, knowing one
of these charges (say $Q_{1,0}$) determines all the other charges.
However, the information contained in these conserved charges
becomes useful if we have a system of multiple (say $n$)
D0-branes. In that case the tachyon on each D0-brane can roll
independently with parameter values $\tl_1$, $\ldots$, $\tl_n$,
and furthermore, there may be arbitrary time delay between the
motion of these tachyons, reflected in the freedom of shifting
$x^0$ for the rolling tachyon backgrounds by arbitrary constants
$c_1$, $\ldots$, $c_n$. The conserved charges $Q_{j,m}$ for such a
system are given by:
 \be \label{eqjmsum}
Q_{j,m}(x^0) = {1\over g_s}\, (-1)^{2m} \, e^{2m x^0} \,
\sum_{l=|m| }^{j-1} b^{l,m}_{j} \sum_{r=1}^n \, e^{2mc_r} \,
\cos^2(\pi\tl_r)  \sin^{2l}(\pi\tl_r)\, .
 \ee
In principle, knowing the infinite number of $Q_{j,m}$'s we can
determine the parameters $\lambda_r$ and $c_r$ for a given
configuration.

\sectiono{Asymptotic Fields Produced by the Rolling Tachyon}
\label{sasrol}

Since the boundary state acts as source for closed string fields,
it produces closed string background $|\Phi\ra$. In the
normalization convention we are using, the equation of motion
determining $|\Phi\ra$ takes the form\cite{0402157}:
 \be \label{eph1}
 2 (Q_B+\bar Q_B)|\Phi\ra = g_s^2 |\BB\ra\, .
 \ee
 The solution to this equation in Siegel gauge is given by:
 \be \label{eph2}
 |\Phi\ra = g_s^2 \, (b_0+\bar b_0) \, \left( 2(L_0+\bar
 L_0)\right)^{-1} |\BB\ra\, .
 \ee
 The right hand side of \refb{eph2} is ambiguous due to the
 existence of zero eigenvalue of $(L_0+\bar L_0)$ in the Minkowski
 space.
 We use the Hartle-Hawking prescription in which we compute
 the result in the Euclidean theory and then analytically continue
 the result to Minkowski theory\cite{0303139,0304192}. As was
 discussed in \cite{0402157}, $|\Phi\ra$
 produced by the $|\BB_1\ra$ part of the boundary state represents
 the closed string radiation produced by the rolling tachyon solution
 whereas the
 closed string background produced by $|\BB_2\ra$ carries
 information about the conserved charges carried by this system.
We shall focus on the
 field produced by the $|\BB_2\ra$ part of the boundary state and
 discuss some subtleties which were overlooked in the analysis of
 \cite{0402157}.

We begin with the general expression for $|\BB_2\ra$:
 \be
\label{eph3}
 |\BB_2\ra = -{i\over g_s}\, {1\over 2\sqrt{\pi}}\,
\sum_{j\ge 1} \sum_{m=1-j}^{j-1} \, f_{j,m}(\tl) \, \int {dP\over
2\pi} \, \sinh(\pi P) \, {\Gamma(-iP)\over \Gamma(iP)}\,
|j,m\ra\ra_X \otimes |P\ra\ra_{liouville} \otimes
|\BB\ra_{ghost}\, .
 \ee
  The part of $|\BB_2\ra$ which involves the
ground state in the ghost sector and primary states in the matter
and the Liouville sector is given by:
 \be \label{eph4}
  -{i\over
g_s}\, {1\over 2 \sqrt{\pi}}\, \sum_{j\ge 1} \sum_{m=1-j}^{j-1} \,
f_{j,m}(\tl) \, \int {dP\over 2\pi} \, \sinh(\pi P) \,
{\Gamma(-iP)\over \Gamma(iP)}\,  |j,m\ra_X \otimes
|P\ra_{liouville} \otimes (c_0+\bar c_0)\, c_1\bar
c_1|0\ra_{ghost}\, .
 \ee
  In the region of large negative $\vp$,
the liouville primary state $|P\ra_{liouville}$ behaves as
 \be \label{eph5}
|P\ra_{liouville} \simeq e^{(2+iP)\vp(0)} |0\ra_{liouville}
-\left( {\Gamma(iP)\over \Gamma(-iP)}\right)^2  e^{(2-iP)\vp(0)}
|0\ra_{liouville}\, .
 \ee
  Furthermore, acting on the state
$|j,m\ra_X\otimes |P\ra_{liouville} \otimes (c_0+\bar c_0)c_1\bar
c_1|0\ra_{ghost}$, the operator $2(L_0+\bar L_0)$ gives
$(P^2+4j^2)$. Thus for large negative $\vp$ the closed string
field configuration produced by the component of the boundary
state given in \refb{eph4} is given by:
 \ben \label{eph6}
|\wc\Phi^{(2)}\ra &=& -{ig_s \over \sqrt{\pi}}\, \sum_{j\ge 1}
\sum_{m=1-j}^{j-1} \, f_{j,m}(\tl) \, \int {dP\over 2\pi} \,
{1\over P^2+4j^2} \, \sinh(\pi P) \, {\Gamma(-iP)\over
\Gamma(iP)}\,  |j,m\ra_X \nonumber \\
&& \otimes \left(
 e^{(2+iP)\vp(0)} |0\ra_{liouville} -\left( {\Gamma(iP)\over
\Gamma(-iP)}\right)^2  e^{(2-iP)\vp(0)} |0\ra_{liouville} \right)
\otimes c_1\bar c_1|0\ra_{ghost} \, . \nonumber \\
 \een
We can evaluate the first term by closing the contour in the lower half
plane and the second term by closing the contour in the upper half plane.
The two contributions are identical and the sum of them
gives:\footnote{Note that the classical background produced by the
D0-brane corresponds to non-normalizable states in the notation of
ref.\cite{seiberg}. The same result is true for string theories
based on $c<1$ minimal models coupled to gravity\cite{0406030}.}
 \be \label{eback1}
|\wc\Phi^{(2)}\ra =  \sqrt \pi \, g_s \sum_{j\ge 1}
\sum_{m=-(j-1)}^{j-1} {1\over ((2j)!)^2} f_{j,m}(\tl) |j,m\ra_{X}
\otimes e^{2(1+j)\vp(0)} |0\ra_{liouville} \otimes c_1\bar c_1
|0\ra_{ghost}\, .
 \ee
This expression agrees with the corresponding expression in
\cite{0402157} up to factors of $g_s$ which was set to 1 in
\cite{0402157}.\footnote{For a recent discussion of why such
backgrounds may be needed to describe a D0-brane from the
viewpoint of the matrix model, see \cite{0401067}.}

Now from eq.\refb{e30} we have
 \be \label{efinterm-}
f_{j,-m}(\tl) = -g_s\sum_{|m|+1\le j'\le j\atop j-j'\in Z} {(2j')!
(2j'-1)!} Q_{j',m} e^{-2mx^0} \, .
 \ee
Eqs.\refb{eback1}, \refb{efinterm-} relate the asymptotic form of
$|\Phi\ra$ produced by the rolling tachyon background to the
conserved charges $Q_{j,m}$ carried by the boundary state. In
section \ref{sadm} we shall derive a general formula relating
asymptotic field configurations to conserved charges.

In ref.\cite{0402157} it was argued that the rest of the
asymptotic field configuration produced by $|\BB_2\ra$, involving
states with oscillator excitations, could be gauged away. This was
done by showing that the boundary state $|\BB_2\ra$ does not
produce any source term in the region of large negative $\vp$.
Eq.\refb{eph1} then shows that the string field $|\Phi\ra$
produced by $|\BB_2\ra$ must be BRST invariant in this region.
(Here we regard the BRST operator as a differential operator in
the $(x,\vp)$ space.) Since the states appearing in expression
\refb{eback1} are the only non-trivial elements of the BRST
cohomology, it then follows that any other contribution to
$|\Phi\ra$ in this region must be BRST trivial and hence can be
gauged away.

There is however a subtlety in this argument which we wish to
discuss here. For this let us briefly recall the argument of
\cite{0402157} for the vanishing of the sources for large negarive
$\vp$. For a D0-brane the Liouville part of the boundary state is
proportional to\footnote{In \refb{ere1} $\vp$ denotes the zero
mode of $\vp$ labelling the space coordinate. We shall use the
same symbol $\vp$ to label the full Liouville field $\vp$ and its
zero mode, but it should be clear from the context which
interpretation is appropriate.}
 \be \label{ere1} \int dP \, \sinh(\pi P) \,
{\Gamma(-iP)\over \Gamma(iP) } |P\ra\ra_{liouville} \, .
 \ee
 If we focus on string field components which
do not involve any oscillator excitation in the Liouville sector,
then for large negative $\vp$ \refb{ere1} gives a source term for
these fields proportional to
 \be \label{ere2}
\int dP \, \sinh(\pi P) \, {\Gamma(-iP)\over \Gamma(iP)} \,
e^{(2+iP)\vp}\,
\, .
 \ee
This can be evaluated by closing the contour in the lower half plane.
Since $\sinh(\pi P) \, {\Gamma(-iP)\over \Gamma(iP)}$ does not have a
pole in the lower half plane, the result vanishes.
Thus $|\BB_2\ra$ does not contribute to any source term for these fields
for large
negative $\vp$.

A higher level state in the Liouville sector, appearing from the
higher level states in $|P\ra\ra_{liouville}$, will typically have
a $P$ dependent coefficient $f(P)$. Thus the source term
associated with such a string field component will be proportional
to
 \be \label{ere3}
\int dP \, \sinh(\pi P) \, f(P) \, {\Gamma(-iP)\over \Gamma(iP)} \,
e^{(2+iP)\vp}\,
\, .
 \ee
As long as $f(P)$ does not have any pole in the lower half $P$
plane, our argument for the vanishing of this integral for large
negative $\vp$ still goes through. However typically $f(P)$ does
have poles, and the residues at the poles correspond precisely to
the null states of the Virasoro algebra. To see an example of this
consider the expansion of $|P\ra\ra_{liouville}$ to states of
level (1,1):
 \be \label{ere4}
|P\ra\ra_{liouville} = \left( 1 + {2\over P^2 + 4} L^\vp_{-1} \bar
L^\vp_{-1}+\ldots\right) |P\ra_{liouville}\, ,
 \ee
where $L^\vp_n$, $\bar L^\vp_n$ denote the Liouville Virasoro
generators, $|P\ra_{liouville} = V_{2+iP}(0)|0\ra_{liouville}$ and
$\ldots$ stand for higher level terms.
 The pole at $P=-2i$ gives a
contribution to \refb{ere1} proportional to
 \be \label{ere5}
 L^\vp_{-1} \bar L^\vp_{-1} e^{4\vp(0)}|0\ra\, ,
 \ee
for large negative $\vp$.
This is a null state of the Liouville Virasoro algebra which does not
vanish when we express the $L^\vp_n$, $\bar L^\vp_n$ in terms of the
oscillators of $\vp$.

The net conclusion from this is that the source term produced by
$|\BB_2\ra$ vanishes for large negative $\vp$ {\it only if we set
the null states to zero}. This in turn shows that the string field
background $|\Phi\ra$ produced by the boundary state in the region
of large negative $\vp$ is BRST invariant only if we set the null
states to zero.\footnote{Operationally this means that we take a
complete basis of states (not necessarily orthonormal) in our
conformal field theory to be the one generated by the action of
$X$ and ghost oscillators and the Liouville Virasoro generators on
primaries of the form $e^{ikX(0)}|0\ra_X \otimes
|P\ra_{liouville}\otimes |0\ra_{ghost}$, and specify a state in
the CFT by specifying its inner product with all members of this
basis. Since a null state is orthogonal to any member of the
basis, it can be regarded as zero.} However, in this case there
are additional elements of the BRST cohomology in the ghost number
two sector, which for large negative $\vp$ take the
form\cite{IMM}:
 \be \label{ere6}
 g_s\, \sum_n n (c_{-n} c_n - \bar
c_{-n} \bar c_n) \, \RR^L_{j-1} \, \RR^R_{j-1} \, e^{2im X(0)}\,
e^{2(1+j)\vp(0)} \, c_1 \bar c_1 |0\ra\, ,
 \ee where
$\RR^L_{j-1}$, $\RR^R_{j-1}$ are the same combination of
oscillators which appear in the construction of
$|\OO^L_{j-1,m}\ra$, $|\OO^R_{j-1, m}\ra$ in eq.\refb{exx5--}.
Thus the background string field $|\Phi\ra$ produced by the
rolling tachyon boundary state will also contain linear
combination of these states. We can in principle calculate the
coefficients of these terms using the analog of \refb{eph6} for
higher level states, but we shall not do this here. For later use
we note however that the state given in \refb{ere6} carries SU(2)
quantum numbers $(j-1,m)$ both on the right and the left sector of
the world-sheet. Thus it must come from terms in $|\BB_2\ra$
carrying SU(2) quantum numbers $(j-1,m)$, and as a result the
coefficient of this term must be proportional to $f_{j-1,m}(\tl)$.

\sectiono{Relation to Conserved Charges in the Matrix Model} \label{s4}

The two dimensional string theory has an equivalent description as a
matrix model\cite{GROMIL,BKZ,GINZIN}. This matrix model in turn can be
mapped to a theory
of free fermions, each of them moving under an inverted harmonic
oscillator Hamiltonian:
 \be \label{e31}
h(q,p) = {1\over 2} (p^2 - q^2) +
{1\over g_s}\, ,
 \ee
where $g_s$ is a parameter which corresponds to the coupling
constant of the closed string theory in the continuum description.
The fermi level is at zero energy; thus all negative energy states
are filled and positive energy states are empty. A single D0-brane
of energy $E$ corresponds to a single fermion excited from the
fermi level to an energy level $E$\cite{0305159}. Thus the
dynamics of a single D0-brane is described by the single particle
Hamiltonian \refb{e31} with an additional constraint:
 \be \label{e32}
h(q,p) \ge 0\, ,
 \ee
due to Pauli exclusion principle. \refb{e31} together with the
constraint \refb{e32} can be regarded as the open string field
theory of a single D0-brane in two dimensional string theory. The
configuration $(q=0, p=0)$ represents the D0-brane with all open
string fields set to zero.

The rolling tachyon solution of the continuum theory parametrized
by $\tl$ corresponds to a classical trajectory in this open string
field theory with energy\cite{0304224,0305159,0305194}
 \be \label{e33}
E = {1\over g_s} \cos^2(\pi
\tl)\, .
 \ee
This gives
 \be \label{e34}
q = -\sqrt{2\over g_s} \,
\sin(\pi\tl) \cosh x^0, \qquad p = -\sqrt{2\over g_s} \, \sin(\pi\tl)
\sinh x^0\, ,
 \ee
where for definiteness we have taken the trajectory to
be in the negative $q$ region. At the classical level and for energy
$E<g_s^{-1}$ there is no tunneling from the negative $q$ to the positive
$q$ region. \refb{e34} may be rewritten as
 \be \label{e36}
(q \pm p) = - \sqrt{2\over g_s} \, \sin(\pi\tl) e^{\pm
x^0}\, .
 \ee
We can now express the conserved charges $Q_{j,m}(x^0)$ of
the continuum theory in terms of matrix model variables\cite{0402157} by
reexpressing \refb{enewqjm} in terms of $q$ and $p$ using \refb{e36}.
This gives
 \be \label{e37a}
Q_{j,m} = \sum_{l=|m| }^{j-1} \left({g_s\over 2}\right)^{l}
b^{l,m}_{j} \, W_{l,m} \, ,
 \ee
 where
 \be \label{ewlm}
 W_{l,m} = h(q, p)\, (q+p)^{l+m}(q-p)^{l-m}\, .
 \ee
The charges $e^{-2mx^0} W_{l,m}$ generate symmetries of the action
corresponding to the Hamiltonian \refb{e31}, and leave the
condition \refb{e32} invariant. Thus \refb{e37a} expresses the
charges in the continuum theory as a linear combination of the
charges in the matrix model description.

Eq.\refb{e37a} can be inverted\cite{0402157} to give
 \be \label{e39a}
  W_{l,m} =  \left({g_s\over 2}\right)^{-l} \,
\sum_{j=|m|+1}^{l+1} c^{l,m}_{j} Q_{j,m}
 \ee
where the coefficients $c^{l,m}_j$ are determined from the
equations:
 \be
\label{e39ba}
 c^{l,m}_{j'} = 0 \quad \hbox{for \quad $j'<|m|+1$
\, or \, $j'>l+1$ \, or \, $l<|m|$} \, ,
 \ee
 \be \label{e39b}
  \sum_{l=|m| }^{j-1} b^{l,m}_{j} c^{l,m}_{j'} = \delta_{jj'} \,
\, \quad \hbox{for \quad $|m|+1\le j'\le j$ } \, .
 \ee
A systematic procedure for solving these equations is to first use
eq.\refb{e39b} for $(m,j')=(j-1,j)$ to determine $c^{j-1,j-1}_j$
for all $j$. Next we use these equations for $(m,j')=(j-2,j)$ and
$(j-2,j-1)$ to determine $c^{j-1,j-2}_j$ and $c^{j-1,j-2}_{j-1}$.
Proceeding this way we can eventually determine all the
coefficients $c^{l,m}_{j'}$ for $|m|+1\le j'\le l+1$.

Note that the relations \refb{e37a} between the conserved charges
$Q_{j,m}$ in the continuum theory and those in the matrix model
have been derived by working with the single fermion excitations
in the range $0\le E \le {1\over g_s}$. However, once these
relations have been derived they hold for all ranges of the
energy. For example, we could extend our expressions for $Q_{j,m}$
is to fermions carrying energy larger than ${1\over g_s}$. The
trajectory associated with a configuration of this type is given
by:
 \be \label{elargeener}
q = -\sqrt{2\over g_s} \, \sinh(\pi\tl) \, \sinh(x^0)\, ,
\qquad
p = -\sqrt{2\over g_s} \, \sinh(\pi\tl) \, \cosh(x^0)\, .
 \ee
The associated $Q_{j,m}$, as computed from \refb{e37a},
\refb{ewlm} are given by:
 \be \label{elarge2}
Q_{j,m}(x^0) = {1\over g_s}\, e^{2m x^0} \,  \cosh^2(\pi\tl)
\sum_{l=|m| }^{j-1} (-1)^{l+m} b^{l,m}_{j} \sinh^{2l}(\pi\tl)\, .
 \ee
These results can also be derived directly from the rolling
tachyon boundary state with energy $> {1\over g_s}$\cite{0203211}.

In section \ref{s5} we shall discuss how \refb{e37a} can be used
to derive constraints on the description of the hole states in the
continuum theory.

\sectiono{Comments on Hole States} \label{s5}

As discussed in section \ref{s4}, the D0-brane in the continuum
description of two dimensional string theory can be identified
with single fermion excitations in the matrix theory. The matrix
theory contains another set of states which are closely related to
the single fermion excitations, -- namely the single hole
excitations. This involves taking a fermion with energy below the
fermi level, and exciting it to an energy just at the fermi level.
The question that arises naturally is: what is the description of
these states in the continuum description of the two dimensional
string theory?

Analysis of the conserved charges carried by a hole state provides
a clue. Since a hole associated with the trajectory $(q(x^0),
p(x^0))$ corresponds to absence of the fermion associated with
this trajectory, the charges $W_{l,m}$ associated with a hole are
given by the negative of the expression given in \refb{ewlm}:
 \be \label{ewhole}
 W^h_{l,m} = -h(q, p) \, (q+p)^{l+m}(q-p)^{l-m} \, .
 \ee
 Hence from \refb{e37a} the charges $Q_{j,m}$ carried by a hole state are
 given by
 \be \label{e37b}
Q^h_{j,m} =  -h(q, p) \, \sum_{l=|m| }^{j-1} \left({g_s\over
2}\right)^{l} b^{l,m}_{j} (q+p)^{l+m}(q-p)^{l-m} \, .
 \ee
In particular, if we take the following negative energy trajectory
 \be \label{enegtraj}
q = -\sqrt{2\over g_s} \, \cosh(\pi\alpha) \, \cosh(x^0)\, ,
\qquad p = -\sqrt{2\over g_s} \, \cosh(\pi\alpha) \, \sinh(x^0)\,
,
 \ee
then the hole state associated with this trajectory carries
 \be \label{eholeqjm}
Q^h_{j,m}(x^0) ={1\over g_s}\, e^{2m x^0} (-1)^{2m}\,
\sinh^2(\pi\alpha) \sum_{l=|m| }^{j-1} b^{l,m}_{j}
\cosh^{2l}(\pi\alpha)\, .
 \ee
Note that $Q^h_{j,m}$ given in \refb{eholeqjm} are related to
$Q_{j,m}$ carried by a fermion, as given in \refb{enewqjm}, by a
formal replacement of $\tl$ by ${1\over 2}+i\alpha$ and an overall
change in sign.

Now suppose the hole state corresponds to some D-brane system in
the continuum theory. Since according to the analysis of section
\ref{s3a} certain terms of the boundary state carry information
about these conserved charges, the charges $Q_{j,m}$ computed from
this boundary state using eqs.\refb{e20}, \refb{edefqjm} must
agree with \refb{eholeqjm}. In particular, these terms must be
given by simply continuing the corresponding terms in the boundary
state of a D0-brane to $\lambda={1\over 2}+i\alpha$, and then
changing the sign of these terms. This is in accordance with the
proposal put forward in \cite{0307195,0307221} where it was
suggested that the complete boundary state of the hole state is
obtained by analytic continuation of $\lambda$ to ${1\over
2}+i\alpha$ followed by a overall change in sign of the boundary
state of a D0-brane. Unfortunately this prescription does not seem
to work for the $|\BB_1\ra$ component of the boundary state. To
see this we note that analytically continuing $-|\BB_1\ra$ to
$\tl={1\over 2}+i\alpha$ amounts to replacing $f(x)$ defined in
\refb{eb3a} by
 \ben \label{efrepl}
\wt f(x) &=& - \left({1\over 1 + \cosh(\pi\alpha)e^{ix}}
+ {1\over 1 + \cosh(\pi\alpha)e^{-ix}}-1\right)
\nonumber \\
&=& {1\over 1 + \hbox{sech}(\pi\alpha)e^{ix}}
+ {1\over 1 + \hbox{sech}(\pi\alpha)e^{-ix}}-1
\, .
 \een
Thus the net effect of this is to replace $\sin(\pi\tl)$ in the
original expression by a $\hbox{sech}(\pi\alpha)$. Since both
$\sin(\pi\tl)$ and $\hbox{sech}(\pi\alpha)$ are less than 1, there
is no qualitative difference between the closed string field
configuration produced by $|\BB_1\ra$ and the new boundary state
$|\wt\BB_1\ra$. In particular it is known that the closed string
field configuration induced by $|\BB_1\ra$ corresponds to a
configuration of the closed string tachyon that describes
precisely a single fermion excitation of the matrix model with a
time delay given by $-\ln\sin(\pi\tl)$\cite{0305159,0402157}. Thus
the closed string field configuration induced by $|\wt\BB_1\ra$
also describes a single fermion excitation of the matrix model
with a time delay proportional to $-\ln\hbox{sech}(\pi\alpha)$ if
we use the same Hartle-Hawking prescription to compute the closed
string radiation from the brane.\footnote{If however we first
compute the closed string radiation from a D0-brane and then
analytically continue the result to $\lambda={1\over 2}+i\alpha$
followed by a change of sign, we get the answer expected of a hole
state\cite{0307195,0307221}. This is due to the fact that the
poles of $f(x)$ crosses the Hartle-Hawking contour as $\tl$ passes
through ${1\over 2}$, and hence the analytic continuation of $\tl$
to ${1\over 2}+i\alpha$ does not commute with integration along
the Hartle-Hawking contour.} In contrast a hole state should be
described by a closed string field configuration of opposite sign,
and by examining the classical trajectory of a hole one can argue
that the corresponding time delay should be given by
$-\ln\cosh(\pi\alpha)$ instead of $-\ln {\rm sech}(\pi\alpha)$.

As a result of these discrepancies, finding the correct expression
for the boundary state describing the hole remains an open
problem. It is possible that the expression for the conserved
charges together with the requirement of world-sheet conformal
invariance determines the form of the boundary state completely.
Unfortunately for a time dependent background the full implication
of world-sheet conformal invariance (the Cardy
conditions\cite{cardy}) is not clear. Trying to understand this
CFT as an analytic continuation from an Euclidean theory might
shed some light on this issue.

Another possible guideline we could try to follow is to try to use
solutions which have a target space interpretation for large
negative $\vp$ and hence are likely to exist in the full string
theory. In this context we note that if we are in the region of
large negative $\vp$ where the potential $e^{2\vp}$ for the
Liouville field is negligible, we can treat $\vp$ as an ordinary
spatial direction with a linear dilaton background $\Phi_D=2\vp$.
In this case we can construct a family of solutions describing the
motion of an ordinary D0-brane in this linear dilaton
background\cite{0405058}.\footnote{I wish to thank A.~Strominger
and J.~Karczmarek for discussion on the issue of interpretation of
ordinary D0-branes in two dimensional string theory.} Such branes
experience a potential proportional to their mass $e^{-\Phi_D}$
and tend to move towards the strong coupling region so as to
minimize their effective mass. Thus these branes are never
stationary. This is indeed one of the characteristics of the hole
states since they involve orbits below the fermi level, and unlike
the $\lambda=0$ D0-brane configuration describing a fermion
sitting at the maximum of the potential, there is no stationary
orbit below the fermi level. The full time dependent classical
solution involves a D0-brane travelling from the strong coupling
to the weak coupling region, reaching a turning point depending on
the total energy it carries, and then going back towards the
strong coupling region. The more the energy the deeper the
D0-brane probes into the weak coupling region. Thus there is no
upper limit to how much energy such a D0-brane can carry.  This
again matches the characteristic of the hole states which do not
have any limit on how much energy they can carry due to the
inverted harmonic oscillator potential being unbounded from below.
The fact that the effective action describing the dynamics of
these holes\cite{0405058} looks very similar to the one used for
describing the dynamics of rolling tachyon\cite{0204143} provides
additional support to this proposed identification.

Although we do not have a complete calculation of the charges
$Q_{j,m}$ carried by these time dependent D0-brane configurations,
for high energy D0-branes which probe well into the weak coupling
region we can make an estimate of $Q_{j,m}$ by computing the
conserved charges at the instant $x^0=0$ when the D0-brane is at
the lowest value of $\vp$. For this suppose the lowest value of
$\vp$ that a D0-brane reaches is $-K$ for some constant $K$. Since
the computation of $Q_{j,m}$ involves computing the disk one point
function of $\phi_{j,m}$ which has a $\vp$ dependence of the form
$e^{2(1-j)\vp}$ for large negative $\vp$, we expect that at
$x^0=0$, $Q_{j,m}$ for this configuration will be of order
 \be \label{eestimate}
 {1\over g_s} \, e^{-2\vp} \, e^{2(1-j)\vp} \bigg|_{\vp = -K}
 \sim {1\over g_s} \, e^{2 j K}\, .
 \ee
 In \refb{eestimate} the first factor of $g_s^{-1}\,
 e^{-2\vp}$ reflects the
 effect of the overall factor of $g_s^{-1}\,
 e^{-\Phi_D}=g_s^{-1}\, e^{-2\vp}$ in the
 D0-brane world-volume action. On the other hand for a high energy
 hole state \refb{eholeqjm} for large
 $\alpha$ gives
 \be \label{eest3}
 Q_{j,m}(x^0=0) \sim {1\over g_s} \, e^{2 j\alpha}\, .
 \ee
 Thus \refb{eestimate} and \refb{eest3} agree under the
 identification $\alpha = K$.

 The boundary state for these hole solutions
have recently been proposed for $\mu=0$\cite{0406173}. In the
euclidean version these branes correspond to the hairpin brane
solutions\cite{0310024,0312168}, which for large negative $\vp$
take the form of a pair of parallel D-branes with Neumann boundary
condition on $\vp$ and Dirichlet boundary condition on $X$. Thus
we would expect that in the $\mu\ne 0$ theory, for large negative
$\vp$ these branes will look like a pair of FZZT
branes\cite{0001012,0009138} with Dirichlet boundary condition on
$X$. This suggests that the CFT description of these branes might
come from beginning with FZZT branes with Dirichlet boundary
condition on $X$(which has $T_{xx}=0$ and hence must represent a
zero energy hole) and then deforming the world-sheet theory by an
appropriate boundary term. It will be interesting to carry out
this construction and see if the corresponding conserved charges
agree with the expected charges carried by the hole.

\sectiono{$\mu\to 0$ Limit} \label{smu}

So far in our analysis we have set $\mu=1$ by a shift of the
Liouville field $\vp$. In this section we shall bring in the
factors of $\mu$ and then take the limit $\mu\to 0$ in order to
compare our results with those given in
\cite{9108004,9109032,9210105,9201056}.\footnote{Alternatively, we
could keep $\mu$ fixed at 1 and take the large imaginary $\tl$
limit so that the total energy ${1\over g_s}\, \cos^2(\pi\tl)$
becomes large compared to ${1\over g_s}$.. In the matrix model
language this will correspond to taking the limit $|p|,|q|>>
{1\over \sqrt{g_s}}$.} For this we need to shift $\vp\to \vp +
{1\over 2}\ln\mu$ so that the cosmological constant term
$e^{2\vp}$ gets transformed to $\mu e^{2\vp}$. {}From
eqs.\refb{exx11}, \refb{exx8} and \refb{exx5} we see that in the
weak coupling region of large negative $\vp$ the gauge
transformation parameter $\Lambda_{j,m}$ has exponential
dependence on $\vp$ of the form $e^{2(1-j)\vp}$. Thus it will pick
up a multiplicative factor of $\mu^{1-j}$ under this shift of
$\vp$. This suggests that we should now use new gauge
transformation parameters $\wh\Lambda_{j,m}$ which are related to
the old parameters by the relation:
 \be \label{emu1}
\wh\Lambda_{j,m} =  \mu^{j-1} \, \Lambda_{j,m}\, ,
 \ee
so that $\wh\Lambda_{j,m}$ do not have any explicit
$\mu$-dependence in the weak coupling region. The corresponding
charges $\wh Q_{j,m}$ are related to $Q_{j,m}$ by a multiplicative
factor of $\mu^{j-1}$:
 \be \label{emu2}
\wh Q_{j,m} = \mu^{j-1} \, Q_{j,m}\, .
 \ee

The shift in $\vp$ also induces a redefinition of the closed
string coupling constant $g_s$. To see this note that before the
shift the kinetic terms for closed string fields are multiplied by
an overall factor of $g_s^{-2}e^{-2\Phi_D}=g_s^{-2} e^{-4\vp}$.
After shifting $\vp$ by ${1\over 2}\ln\mu$ this factor becomes
$g_s^{-2}\mu^{-2} e^{-4\vp}$. This suggests that we define a new
closed string coupling constant
 \be \label{edefgs}
\wh g_s = g_s \, \mu\, .
 \ee

We shall now express $\wh Q_{j,m}$ carried by a single D0-brane in
terms of the matrix model variables. Replacing the ${1\over g_s}$
factor in the matrix model by ${\mu\over \wh g_s}$ in
eqs.\refb{e37a}, \refb{ewlm} gives\footnote{Since in the $\mu\to
0$ limit the potential barrier at the fermi level between the
negative $q$ side and the positive $q$ side disappears, we expect
that this limit gives a sensible theory only when negative energy
levels on both sides of the barrier are filled. Thus strictly
speaking the analysis of this section and of section \ref{sblack}
will be sensible only when we repeat this in the type 0B string
theory\cite{0307083,0307195}. However since our analysis is based
on the study of conserved charges and does not involve the details
of the dynamics, we expect that the results obtained from our
analysis will survive in the full theory where negative energy
states on both sides of the barrier are filled, even though the
system that we are studying has negative energy states filled only
on one side of the barrier.}
 \be \label{e37aa}
Q_{j,m} = h(q, p)\, \sum_{l=|m| }^{j-1} \left({ \wh g_s\over
2\mu}\right)^{l} b^{l,m}_{j} (q+p)^{l+m}(q-p)^{l-m}
 \ee
with
 \be \label{enewh}
h(q, p) = {p^2\over 2}-{q^2\over 2}+{\mu\over
\wh g_s}\, .
 \ee
Hence from \refb{emu2}
 \be \label{emu3}
\wh Q_{j,m} = \mu^{j-1} \, \left({p^2\over 2}-{q^2\over 2}+{ \mu
\over \wh g_s}\right) \, \sum_{l=|m| }^{j-1} \left({\wh g_s\over
2\mu}\right)^{l} b^{l,m}_{j} (q+p)^{l+m}(q-p)^{l-m} \, .
 \ee

Let us now consider the $\mu\to 0$ limit of \refb{emu3} keeping
$\wh g_s$, $p$ and $q$ fixed. In this limit only the $l=(j-1)$
term contributes to the sum, and we get
 \be \label{emu4}
\wh Q_{j,m} =  -{1\over 2} \, \left({ \wh g_s\over 2}\right)^{
j-1} b^{j-1,m}_{j} (q+p)^{j+m}(q-p)^{j-m}\, .
 \ee
{}From \refb{e38a} and the fact that $\alpha^{j-|m|-1}_{j-1,-m}$
vanishes since $\alpha^s_{j',m}$ vanishes for $s>j'-|m|-1$, we
get:
 \be \label{emore2}
b^{j-1,m}_{j} = -{1\over (2j)!(2j-1)!}\, \alpha^{j-|m|-1}_{j,-m}
\, .
 \ee
 An explicit expression for $\alpha^{s}_{j,m}$ has been given in
eq.\refb{efjm1}. From this one can show that
 \be \label{ealjmsp}
\alpha^{j-|m|-1}_{j,m} = -{(2j)!\over (j+m)! (j-m)!}\, .
 \ee
The identity required for proving this is obtained by comparing
the coefficients of $x^{j-m}y^{j+m}$ in the equation:
 \be \label{etrivial}
(x+y)^{2j} = (x+y)^{j+m} (x+y)^{j-m}\, ,
 \ee
by expressing each of the three factors in a binomial expansion.
Using eqs.\refb{emore2} and  \refb{ealjmsp}, eq.\refb{emu4} may be
expressed as
 \ben \label{emore3}
\wh Q_{j,m} &=& {1\over 2}\left({\wh g_s\over 2}\right)^{ j-1}
{1\over (2j)!(2j-1)!}\, \alpha^{j-|m|-1}_{j,-m} \,
(q+p)^{j+m}(q-p)^{j-m} \nonumber \\
&=& -{1\over 2} \, {1\over (2j-1)! (j+m)! (j-m)!} \, \left({\wh
g_s\over 2}\right)^{ j-1} \, (q+p)^{j+m}(q-p)^{j-m} \, .
 \een
Up to an overall normalization constant this expression for $\wh
Q_{j,m}$ agrees with the results of
\cite{9108004,9109032,9210105,9201056} based on the analysis of
the algebra of symmetries generated by these transformations. Some
issues regarding normalization factors have been discussed in
appendix \ref{appb}.

It is also instructive to consider the $\mu\to 0$ limit of the
discrete state closed string field configuration produced by the
$|\BB_2\ra$ component of the boundary state. After the shift of
$\vp$ by ${1\over 2}\ln\mu$, \refb{eback1} takes the form:
 \be \label{eback2}
|\wc\Phi^{(2)}\ra = \wh g_s \sum_j \sum_{m=-(j-1)}^{j-1} {\sqrt
\pi\over ((2j)!)^2} \mu^j \, f_{j,m}(\tl) |j,m\ra_{X} \otimes
e^{2(1+j)\vp(0)} |0\ra_{liouville} \otimes c_1\bar c_1
|0\ra_{ghost}\, .
 \ee
The relation between $f_{j,m}$ and $Q_{j',m'}$ has been given in
\refb{efinterm-}. Using \refb{emu2} and \refb{edefgs} we can
rewrite this as
 \be \label{efinterm}
f_{j,-m}(\tl) = -\wh g_s \sum_{1\le j'\le j\atop j-j'\in Z}
{(2j')! (2j'-1)! } \mu^{-j'} \, \left(\wh Q_{j',m}
e^{-2mx^0}\right) \, .
 \ee
Substituting this into \refb{eback2} we see that only the $j'= j$
term contributes in the $\mu\to 0$ limit, and gives:
 \be \label{eback9}
|\wc\Phi^{(2)}\ra =  - (\wh g_s)^2 \, \sum_j \sum_{m=-(j-1)}^{j-1}
{\sqrt \pi\over 2j} \, \left( e^{2mx^0} \, \wh Q_{j,-m}\right) \,
|j,m\ra_{X} \otimes e^{2(1+j)\vp(0)} |0\ra_{liouville} \otimes
c_1\bar c_1 |0\ra_{ghost}\, .
 \ee
 Thus $|\wc\Phi^{(2)}\ra$ has a finite non-zero $\mu\to 0$ limit.

 As argued in section \ref{sasrol}, $|\BB_2\ra$ also produces
 additional closed string background involving
 states of the form given in \refb{ere6}, with coefficient proportional
 to $f_{j-1,m}(\tl)$. When we express $f_{j-1,m}(\tl)$ in terms of
 $\wh Q_{j',m'}$ by replacing $j$ by $j-1$ in \refb{efinterm}, the
 leading contribution is of order $\mu^{-(j-1)}$. On the other
 hand the shift of
 $\vp$ by ${1\over 2}\ln\mu$ in \refb{ere6} and expressing
 the overall factor of $g_s$ in terms of
 $\wh g_s$ produces the same factor of $\mu^j$
 as in \refb{eback2}. Thus in the $\mu\to 0$ limit the
 contribution from this term vanishes. This shows that \refb{eback9}
 is the only contribution left in this limit. This is consistent
 with the fact that for $\mu=0$ we can choose to use the
 oscillator description of the field $\vp$, and in that case the
 additional states appearing in \refb{ere6}
 are not BRST invariant as $(Q_B+\bar Q_B)$
 acting on these states do not vanish when we express all
 states in terms of $\vp$ oscillators. Thus had the coefficients
 of the additional states not vanished in the $\mu\to 0$ limit, we
 shall be left with a field configuration that is not on-shell
 asymptotically.

For $\mu=0$ the string coupling $\wh g_s$ is an irrelevant
constant. This is obvious in the matrix model since $h(q,p)$ given
in \refb{enewh} is independent of $\wh g_s$ in this limit. In the
continuum description we see this by noting that a shift $\vp\to
\vp -{1\over 2}\ln\wh g_s$ removes the $\wh g_s$ from the
pre-factor $\wh g_s^{-2} e^{-4\vp}$ in the closed string action. A
consistency check of eq.\refb{eback9} will be that after this
shift of $\vp$ the relation between $|\wc\Phi^{(2)}\ra$ and the
matrix model variables $q, p$ should be independent of $\wh g_s$.
Is this true? Using \refb{emore3} we see that this shift in $\vp$
changes \refb{eback9} to
 \ben \label{ebackext}
|\wc\Phi^{(2)}\ra &=&  \sqrt{\pi} \, \sum_j \sum_{m=-(j-1)}^{j-1}
\, 2^{-j} \, {1\over (2j)! (j+m)! (j-m)!} \, e^{2m x^0} \,
(q+p)^{j-m} \,
(q-p)^{j+m} \nonumber \\
&& |j,m\ra_{X} \otimes e^{2(1+j)\vp(0)} |0\ra_{liouville} \otimes
c_1\bar c_1 |0\ra_{ghost}\, .
 \een
As required, this equation has no $\wh g_s$ dependence.

\sectiono{Conserved Charges from Asymptotic String Field Configurations}
\label{sadm}

So far in our analysis we have discussed how to compute the
conserved charges $\wh Q_{j,m}$ for a D-brane system in terms of
its boundary state. As discussed in section \ref{sasrol}, since
the boundary state of a D-brane acts as sources for various closed
string fields, it produces closed string field configuration. We
should expect that the asymptotic field configuration also
contains information about the conserved charges of the system,
just as long range electric field and gravitational field contain
information about the charge and mass of a configuration in
Maxwell-Einstein theory. It will be useful to find an expression
for the conserved charges in terms of the asymptotic field
configuration, since such a formula will have a more general
validity even for configurations which are not D-branes. This is
what we shall do now. Our strategy will be to first find the
relation between the conserved charges and asymptotic field
configurations for a D-brane system, and then treat this as a
universal relation that holds for any configuration in this
theory. During this analysis we shall work with general $\mu$ from
the beginning and call the string coupling constant $\wh g_s$, so
that after inverse Wick rotation the conserved charges obtained
from this analysis directly gives the charges $\wh Q_{j,m}$.

Our strategy will be to formally manipulate the expression for the
conserved charge carried by a D-brane to write it as an integral
of a total derivative with respect to the Liouville coordinate
$\vp$, so that we can express it as a boundary term evaluated at
large negative $\vp$. We begin with the expression \refb{e20} of
the conserved charge $F_{j,m}(x)$. In this expression
$|\phi_{j,m}(p)\ra$ is built on a fixed Liouville primary
$V_{2(1-j)}(0)|0\ra_{liouville}$. Let us consider a general family
of states $|\phi_{j,m}(p,q)\ra$ built on the Liouville primary
$V_{2+iq}(0)|0\ra_{liouville}$ such that
 \be \label{eadm9}
|\phi_{j,m}(p,q=2ij)\ra = |\phi_{j,m}(p)\ra\, ,
 \ee
and define:
 \be \label{eadm11}
F_{j,m}(x,q) = \int {dp\over 2\pi} \, e^{-i
p x} \la \BB | (c_0-\bar c_0)|\phi_{j,m}(p,q)  \ra\, .
 \ee
Clearly $F_{j,m}(x, q=2ij)$ gives us the conserved charge
$F_{j,m}(x)$ defined in \refb{e20}. Our goal will be to express
$F_{j,m}(x,q)$ in terms of the closed string field $|\Phi\ra$
produced by the boundary state.  For this we note that since
$|\phi_{j,m}(p)\ra$ is BRST invariant, $(Q_B+\bar
Q_B)|\phi_{j,m}(p,q)\ra$ must vanish at $q=2ij$. This allows us to
define a new family of states $|\Omega_{j,m}(p,q)\ra$ through the
relations:
 \be \label{eadm10}
  (Q_B+\bar Q_B) |\phi_{j,m}(p,q)\ra =
(q - 2ij) |\Omega_{j,m}(p,q)\ra\, . \ee
 Next we note that the linearized
closed string field equation \refb{eph1} with $g_s$ replaced be
$\wh g_s$ takes the form:
 \be \label{eadm12}
- 2 \la\Phi |(Q_B+\bar Q_B) = \wh g_s^2\, \la
\BB|\, .
 \ee
 Using \refb{eadm10}, \refb{eadm12} and the fact that
neither $|\phi_{j,m}(p,q)\ra$ nor the closed string field
$|\Phi\ra$ carries a $(c_0-\bar c_0)$ zero mode,  we can now
express \refb{eadm11} as
 \be \label{eadm13}
F_{j,m}(x,q) = (q - 2ij) G_{j,m}(x, q)\, ,
 \ee
where
 \be \label{eadm14}
G_{j,m}(x,q) = 2 \, \wh g_s^{-2}\, \int {dp\over 2\pi} \, e^{-i
p x}  \la \Phi |(c_0-\bar c_0)|\Omega_{j,m}(p,q)\ra\, .
 \ee
If we define:
 \be \label{eadm15b}
\wt G_{j,m}(x,\vp) =  \int {dq\over 2\pi} \, e^{(2-iq)\vp}
G_{j,m}(x,q)\, ,
 \ee
then \refb{eadm13} gives
 \be \label{eadm17}
F_{j,m}(x,q) = i \int \, d\vp e^{i(q-2ij)\vp} \, \p_\vp \left(
e^{-2(1+j)\vp}\wt
G_{j,m}(x,\vp) \right)\, .
 \ee
Thus
 \be \label{eadm18}
F_{j,m}(x) = F_{j,m}(x, q=2ij) =  i \int \, d\vp \,
\p_\vp \left(
e^{-2(1+j)\vp}\wt
G_{j,m}(x,\vp) \right)\, .
 \ee
The right hand side of \refb{eadm18} is a total derivative in the
Liouville coordinate. Thus we can identify the conserved charge
$F_{j,m}(x)$ as the boundary value of $ -i e^{-2(1+j)\vp}\wt
G_{j,m}(x,\vp)$ as $\vp\to -\infty$:
 \ben \label{eadm19}
F_{j,m}(x) &=& -i\lim_{\vp\to -\infty} \left(e^{-2(1+j)\vp}\wt
G_{j,m}(x,\vp)\right)\nonumber \\
&=& -2i\, \wh g_s^{-2}\, \lim_{\vp\to -\infty} \,
\left(e^{-2(1+j)\vp} \, \int{dp\over 2\pi} \, e^{-ipx} \, \left[
\int {dq\over 2\pi} \, e^{(2-iq)\vp} \, \la\Phi|(c_0-\bar c_0) |
\Omega_{j,m}(p,q)\ra
\right] \right) \, . \nonumber \\
 \een
In this expression the term inside the square bracket may be
interpreted as the $\vp$-space wave-function of an appropriate
component of the state $|\Phi\ra$ at large $\vp$. Thus
\refb{eadm19} after inverse Wick rotation gives the desired
expression for the charges $\wh Q_{j,m}$ in terms of the
asymptotic closed string field configuration. Since this relation
expresses $\wh Q_{j,m}$ in terms of field configurations in the
weak coupling region, we can use this to define $\wh Q_{j,m}$ even
in the $\mu\to 0$ limit.

There are clearly infinite number of ways of defining the
 family of states
 $|\phi_{j,m}(p,q)\ra$ satisfying the requirement that $|\phi_{j,m}(p,q=2ij)
 \ra = |\phi_{j,m}(p)\ra$. This induces an ambiguity in the
 definition of $|\Omega_{j,m}(p,q)\ra$ in the form of the freedom
 of adding a BRST exact contribution.  However
as long as the asymptotic closed string field configuration is
on-shell, the freedom of adding a BRST exact contribution to
$|\Omega_{j,m}\ra$ does not affect the expression \refb{eadm19}
for the conserved charge since a BRST exact state has vanishing
inner product with an on-shell state.

We shall now illustrate this construction for the $\mu=0$ case
that will be of interest to us in section \ref{sblack}. In this
case we can regard $\vp$ as a scalar field with background charge
and use the oscillators of $\vp$ to label states. Let us suppose
$|\Omega_{j,m}(p,q)\ra$ has the form
 \be \label{eadm20b}
|\Omega_{j,m}(p,q)\ra = \wh\Omega_{j,m}(p,q)
e^{ipX(0)}|0\ra_X \otimes e^{(2+iq)\vp(0)}|0\ra_{liouville}
 \otimes
|0\ra_{ghost}  \, ,
 \ee
 for some combination $\wh\Omega_{j,m}(p,q)$ of ghost
oscillators and non-zero mode
 $X$ and
$\vp$ oscillators. Then if the asymptotic closed string field
$|\Phi\ra$ has a term of the form:
\be \label{eadn1}
\wh \Phi\, e^{ip'X(0)}|0\ra_X \otimes
e^{(2+iq')\vp(0)}|0\ra_{liouville}
 \otimes
|0\ra_{ghost}  \, ,
 \ee
for some combination $\wh\Phi$ of  ghost
oscillators and non-zero mode
 $X$ and
$\vp$ oscillators, then the contribution to the right hand side of
\refb{eadm19} from this term in $|\Phi\ra$ is given by:
 \be \label{econtfjm}
  -2i \wh g_s^{-2} \, \lim_{\vp\to -\infty} \,
\left(e^{-2(1+j)\vp} \la\la 0|\wh\Phi^T \, (c_0-\bar c_0) \, \wh
\Omega_{j,m}(-p',-q') |0\ra\ra e^{ip'x} e^{(2+iq')\vp}\right)\, .
 \ee
 Here $\wh\Phi^T$ denotes the BPZ conjugate of
 $\wh\Phi$, and $\la\la
0|\cdot|0\ra\ra$
denotes matrix element involving only the ghost oscillators and
the non-zero mode oscillators of the $X$ and $\vp$ fields.

Since conservation laws imply that $F_{j,m}(x)$ must have the form
$e^{-2im x}$ we see from \refb{econtfjm} that the relevant component
of  $|\Phi\ra$ that contributes to $F_{j,m}(x)$ must have
$p'=-2m$.
As a result in \refb{econtfjm} we can replace $-p'$
in the argument of $\wh\Omega_{j,m}$ by $2m$.
Thus the computation of the value of the conserved charge requires
$|\Omega_{j,m}(p,q)\ra$ for $p=2m$. This
computation can be simplified by the
following
observation.
  Since for $\mu=0$ the only non-trivial elements of BRST
cohomology at ghost number two and Liouville momentum $q=2ij$ are
the states $|Y^L_{j,m}\ra\times |Y^R_{j,m}\ra$, the rest of the
contributions in \refb{ephijmp} are BRST trivial and can be
ignored. This allows us to choose
 \be \label{eklo1}
|\phi_{j,m}(p=2m, q)\ra = |j,m\ra_X \otimes
e^{(2+iq)\vp(0)}|0\ra_{liouville} \otimes c_1\bar c_1|0\ra_{ghost}
 \ee
 and hence from \refb{eadm10}
\be \label{eklo2}
 |\Omega_{j,m}(p=2m, q)\ra ={1\over 4} \, (q+2ij) |j,m\ra_X \otimes
e^{(2+iq)\vp(0)}|0\ra_{liouville} \otimes (c_0+\bar c_0) \,
c_1\bar c_1|0\ra_{ghost} \, .
 \ee
Comparing this with \refb{eadm20b} and using \refb{e13} we see
that
 \be \label{eklo3}
 \wh\Omega_{j,m}(p=2m, q) = {1\over 4}(q+2\, i\, j)\, \PP^L_{j,m} \,
 \PP^R_{j,m} \, (c_0+\bar c_0) \,
c_1\bar c_1 \, . \ee
 \refb{econtfjm} together with the
orthonormality relations \refb{e14} now shows that
 the contribution to $F_{j,m}$ from a $|\Phi\ra$ that is
 on-shell asymptotically comes from the coefficient of the BRST
 cohomology element:
 \be \label{eklo31}
 |j,-m\ra_X \otimes e^{2(1+j)\vp(0)}|0\ra_{liouville} \otimes
 c_1\bar c_1|0\ra_{ghost}\, ,
 \ee
 in $|\Phi\ra$.

We shall now test \refb{econtfjm} by computing the charges from
the asymptotic field configuration given in \refb{eback9} (after
making the replacement $x^0\to -ix$). The only term in $|\Phi\ra$
that contributes to $F_{j,m}(x)$ is the term:
 \be \label{eterm1}
  -
(\wh g_s)^2 \, {\sqrt \pi\over 2j} \, \left( e^{2imx} \,  \wh
Q_{j,m} \right) \, |j,-m\ra_{X} \otimes e^{2(1+j)\vp(0)}
|0\ra_{liouville} \otimes c_1\bar c_1 |0\ra_{ghost}\, .
 \ee
Comparison with \refb{eadn1} shows that
for this term
 \be \label{eterm2}
\wh \Phi = - (\wh g_s)^2 \, {\sqrt \pi\over 2j} \, \left( e^{2imx}
\,  \wh Q_{j,m} \right) \, \PP^L_{j,-m} \,
 \PP^R_{j,-m} \,
c_1\bar c_1 \, , \qquad p'=-2m, \qquad q'=-2ij \, .
 \ee
Substituting \refb{eklo3} and \refb{eterm2} into \refb{econtfjm},
and using the normalization condition \refb{e14} we get
 \ben \label{eterm3}
 F_{j,m}(x) &=& -2i \wh g_s^{-2} \, \lim_{\vp\to -\infty} \,
\left(e^{-2(1+j)\vp} (- (\wh g_s)^2) \, {\sqrt \pi\over 2j} \,
\left( e^{2imx} \,  \wh Q_{j,m}\right) \, (i j) \, (2) \,
e^{-2imx} \, e^{2(1+j)\vp} \right) \nonumber \\
&=& -2\sqrt \pi \, \wh Q_{j,m}\, .
 \een
This agrees with the relation between $F_{j,m}$ and $Q_{j,m}$
introduced in \refb{edefqjm}. One factor of $(2)$ in \refb{eterm3}
comes from the ghost correlator.

Of course in this case this agreement is not a surprise since we have
defined $Q_{j,m}$ through eq.\refb{edefqjm}. However we can take this
agreement as a test of our equations \refb{eadm19},
\refb{econtfjm} relating the
charges $\wh Q_{j,m}$ to the asymptotic field configuration. We shall use
these relations in section \ref{sblack} to compute the charges $\wh
Q_{j,m}$ carried by the black hole.

\sectiono{Two Dimensional Black Holes} \label{sblack}

We can use the results of section \ref{sadm} to calculate the
charges $\wh Q_{j,m}$ for any string field configuration in two
dimensional string theory, and then try to identify a
corresponding matrix model state that carries the same set of
conserved charges. One particular configuration that is of
interest is the two dimensional black hole solution\cite{MSW,WB}
in the $\mu\to 0$ limit. In this section we shall try to compute
the conserved charges $\wh Q_{j,m}$ carried by the black hole and
then try to identify an appropriate configuration in the matrix
model that carries such charges.\footnote{For previous attempts at
identifying the black hole in the matrix model see
\cite{0101011,9210107,marsha,9210120,9303116,9304072,9305109}. It
is not completely clear how to compare our results for $\mu=0$
Lorentzian black hole with the results of these papers.} Since for
$\mu=0$ we can remove the string coupling constant $\wh g_s$ by a
shift of $\vp$, we shall set $\wh g_s=1$ in this analysis.

In order to calculate the conserved charges carried by the black
hole we need to find the closed string field configuration
associated with the black hole for large negative $\vp$. Thus we
need to be able to represent the black hole as a classical
solution in string field theory. We begin by writing down the
solution in the effective field theory in the Schwarzschild like
coordinates\cite{MSW,WB}:
 \be \label{esch}
G_{00} = -(1-ae^{4\vp}), \quad G_{\vp\vp}= (1-ae^{4\vp})^{-1},
\quad \Phi_D = 2\vp\, ,
 \ee
 where $G_{\mu\nu}$ denotes the string metric, $\Phi_D$ denotes
 the dilaton, and $a$ is a parameter related to the mass of the
 black hole. This satisfies the equations of motion derived from
 the action of two dimensional dilaton gravity:
 \be \label{esch1a}
 \SSS = \int dx^0 \, d\vp \, e^{-2\Phi_D} \, \sqrt{-\det G} \,
(R_G + 4 G^{\mu\nu} \p_\mu\Phi_D \p_\nu \Phi_D + 16)\, .
 \ee
 Defining $h_{\mu\nu}$ and $\phi_D$ as the deviation
 of the metric and the dilaton
 from the flat Minkowski metric and linear dilaton background:
  \be \label{esch2}
  h_{\mu\nu} = G_{\mu\nu} - \eta_{\mu\nu}\, , \quad \phi_D = \Phi_D
  - 2\vp\, ,
  \ee
  we see that for large negative $\vp$, \refb{esch} takes the
  form:
  \be \label{esch3}
  h_{00} = ae^{4\vp}, \quad h_{\vp\vp} \simeq ae^{4\vp},
  \quad \phi_D=0\, .
  \ee
  These give a solution of the linearized equations of motion
  derived from the action \refb{esch1a}
  around the linear dilaton background.
  The complete solution \refb{esch} can now be recovered as a power series
expansion in $e^{4\vp}$
by beginning with \refb{esch3} and then iteratively solving the full
equations of
motion derived from \refb{esch1a}. In particular $r$ iterations will
be needed to recover the term of
order $e^{4r\vp}$ in the expansion of the solution \refb{esch} in
powers of $e^{4\vp}$.

We shall now use this construction as a guideline to construct the
black hole solution in closed string field theory\cite{MMS}. We
begin with the observation that up to gauge transformation, the
solution \refb{esch3} corresponds to switching on a closed string
background proportional to
 \be \label{eblackas}
a \, |j=1, m=0\ra_{X}\otimes e^{4\vp(0)}|0\ra_{liouville} \otimes
c_1\bar c_1|0\ra_{ghost}\, .
 \ee
 This, being a BRST invariant state, is clearly a solution of the
 linearized equations of motion of closed string field theory.
 We can now generate the full solution of the closed string field
 equations by beginning with \refb{eblackas} and then
 solving the closed string field equations
iteratively.\footnote{For Euclidean black hole it has been
suggested that the solution is obtained by deforming a flat linear
dilaton background by a non-normalizable operator carrying
non-zero winding number along the $X$
direction\cite{0101011,mald}. The relationship between our
approach for generating the solution in the Minkowski theory and
that of  \cite{0101011,mald} is not clear. We note however that
even a conventional D-brane wrapping a compact direction carries
winding charge along that direction. But once we go to the
universal cover of the circle describing a non-compact direction,
the winding charge disappears from the boundary state describing
the D-brane.} Our analysis is facilitated by the fact that the
initial configuration \refb{eblackas} is independent of the time
coordinate $x^0$. Thus we can work in a restricted subsector of
closed string field theory involving time independent fields ({\it
i.e.} fields carrying zero $X^0$ momentum). This restricted string
field theory action has an $SU(2)_L\times SU(2)_R$ symmetry
inherited from the $SU(2)_L\times SU(2)_R$ symmetry of the
euclidean theory at the self-dual radius. In particular the string
field interaction terms can couple a field of $SU(2)_L$ (or
$SU(2)_R$) quantum number $j$ to a set of fields carrying SU(2)
quantum numbers $j_1, \ldots j_r$ only if $j\le j_1+\ldots j_r$.

It is now clear that the order $e^{4r\vp}$ term, obtained by
iteration of the string field equations with \refb{eblackas} as
the starting point, must be a linear combination of terms of the
form:
 \be \label{eblackas2}
\KK_r \, \left(|j', m=0\ra_L \times |j'', m=0\ra_R\right) \otimes
e^{4r\vp(0)}|0\ra_{liouville} \otimes c_1\bar c_1|0\ra_{ghost}\, ,
\quad j', j''\le r\, ,
 \ee
where $\KK_r$ is some combination of $X$ and Liouville Virasoro
generators and ghost oscillators. Comparing this with
\refb{eklo31} we see that in order that it contributes to
$F_{j,m}(x)$ we must have $j=2r-1$ so as to match the power of
$e^\vp$ in \refb{eklo31} and \refb{eblackas2}. On the other hand
comparing the SU(2) quantum numbers we see that in order that
\refb{eblackas2} contributes to $F_{j,m}(x)$, $j'$ and $j''$ must
be equal to $j$. This gives $j\le r$. This gives the restriction:
 \be \label{erestr} 2r - 1\le r, \qquad {\i.e.}
\qquad r\le 1\, .
 \ee
  Thus we see that all contributions to $F_{j,m}(x)$ vanish for
  $j>1$.
   After inverse Wick rotation this gives:
 \be \label{eqblack}
\wh Q_{j,m}/\wh Q_{1,0} = 0\,  \quad \hbox{for $j>1$}\, .
 \ee

We can now look for a configuration in the matrix model that
carries the same quantum numbers. We focus on the quantum numbers
$\wh Q_{j,0}$. \refb{emore3} shows that any single fermion state
of energy $E$ will carry non-zero $\wh Q_{j,0}$ for all $j$, and
hence cannot represent a black hole. The situation does not
improve much by superposing a finite number of fermion / hole
states of different energies.\footnote{We are working under the
assumption that the Lorentzian black hole can be described within
the free fermion description of the matrix model, and does not,
for example, involve U(N) non-singlet sector of the matrix
model\cite{0101011}.} To see this note that if we have fermions of
energy $E_1$, $E_2$, $\ldots$ and holes of energies $\eps_1$,
$\eps_2$, $\ldots$, so that the total energy of the system is
given by
 \be \label{eblaen}
\sum_k E_k + \sum_r \eps_r\, ,
 \ee
then $\wh Q_{j,0}$ associated with this state, as computed from
\refb{emore3}, is proportional to
 \be \label{ebladis-}
 \left( \sum_r (\eps_r)^j -\sum_k (-E_k)^j
\right) \, .
 \ee
For odd $j$ all the contributions add and hence it seems
impossible to adjust the $ E_k$'s and the $\eps_r$'s so that the
total $\wh Q_{1,0}$ is finite or large while all other $\wh
Q_{j,0}/\wh Q_{1,0}$ ratios vanish. The only way to avoid this
situation will be to take a superposition of many single fermion
and hole states, each with energy $E_k$ or $\eps_r$ close to zero,
so that the total energy adds up to the mass $M$ of the black
hole.\footnote{Since these results were first reported in the
Strings 2004 conference in Paris\cite{asparis}, ref.\cite{0407136}
has made similar observations.} If there are $N$ fermions, and we
take each of $E_k$ and $\eps_r$ to be of order $M/N$, then the
$\wh Q_{j,m}$ associated with this configuration is of order
 \be \label{eback10}
N (M/N)^j \sim M^j N^{1-j}\, .
 \ee
Thus in the $N\to\infty$ limit this vanishes for all $j>1$, as is
required for a black hole.

This analysis suggests that the black hole of two dimensional
string theory can be thought of as a collection of a large number
of fermions and holes, each with a very small energy, so that the
total energy adds up to the mass $M$ of the black hole. This is
consistent with the fact that classically for $\mu=0$ only a pulse
of very small height can stay in the strong coupling region near
the top of the potential for a long period.

If this is really the correct description of the black hole then
it gives rise to an apparent puzzle. Since a black hole consists
of a large number of fermions and holes close to zero energy, a
finite energy fermion or hole will not feel the effect of the
black hole in the classical limit. If we identify these finite
energy excitations with D0-branes as in the $\mu\ne 0$ case, then
this will imply that a classical D0-brane in this string theory
should not be able to distinguish between the black hole geometry
and the usual linear dilaton background. Can this be true?
D0-branes in two dimensional black hole geometry have been
analyzed from various viewpoints in
\cite{0105038,0310024,0406017,yogendran}, and we can borrow these
results to make this comparison. First let us note that the usual
linear dilaton background of two dimensional string theory is
given by
 \be \label{ebc1}
ds^2 \equiv G_{\mu\nu} dx^\mu dx^\nu = -(dx^0)^2 + d\vp^2, \qquad
\Phi_D = 2\vp\, ,
 \ee
where $G_{\mu\nu}$ denotes the string metric, $\Phi_D$ is the
dilaton, and $x^0$ and $x^1\equiv\vp$ are the coordinates of the
two dimensional space. In this background the world-line theory of
a D0-brane is given by the action:\footnote{Note that according to
the proposal made at the end of section \ref{s5} these D0-branes
describe hole states rather than fermion states, but this does not
affect our argument.}$^,$\footnote{Naively we would expect that in
this case the D0-brane, besides containing the mode that allows it
to move along the Liouville direction, will also have a tachyonic
mode. From the description of the D0-brane as a single fermion in
the matrix model, we do not have an obvious identification of this
mode. We note however that in the presence of a linear dilaton
background the D0-brane accelerates, and hence it is not
completely clear if the tachyonic mode is really present even in
the continuum description. On the other hand, due to this
acceleration it is also not clear if the Dirac-Born-Infeld action
provides a reliable description of the D0-brane dynamics. We shall
nevertheless proceed with this action in the absence of a better
alternative.}
 \be \label{ebc2}
-\int d\tau e^{-\Phi_D} \, \sqrt{-G_{\mu\nu}
\p_\tau X^\mu \p_\tau X^\nu} = -\int d\tau \,
\sqrt{-e^{-2\Phi_D} G_{\mu\nu} \p_\tau X^\mu \p_\tau X^\nu} \, ,
 \ee
where $\tau$ is a parameter labelling the D0-brane world-line.
Thus effectively the motion of the D0-brane is described by a
particle moving under the metric\cite{0105038}
 \be \label{ebc3}
ds_{D0}^2 \equiv e^{-2\Phi_D} G_{\mu\nu} dx^\mu dx^\nu = e^{-4\vp}
(-(dx^0)^2 + d\vp^2)\, .
 \ee
On the other hand the black hole background in the Schwarzschild-like
coordinate system is given by\cite{MSW,WB}:
 \be \label{ebc4}
\wt{ds}^2 = - (1 - a e^{4\vp}) (dx^0)^2 + (1 - a e^{4\vp})^{-1}
d\vp^2, \qquad \Phi_D = 2\vp\, ,
 \ee
where $\wt{ds}$ denotes the line element in the black hole background
measured in the string
metric.
The D0-brane moving in this background sees an effective metric
 \be \label{ebc5}
\wt{ds}_{D0}^2 \equiv e^{-2\Phi_D} \, \wt{ds}^2 = e^{-4\vp} \,
\left[- (1 - a e^{4\vp}) (dx^0)^2 + (1 - a e^{4\vp})^{-1}
d\vp^2\right]\, .
 \ee
If we define
 \be \label{ebc6}
w = -{1\over 4} \ln\left( e^{-4\vp} -a\right)\, ,
 \ee
then in this new coordinate system \refb{ebc5} takes the form:
 \be \label{ebc7}
\wt{ds}_{D0}^2 = e^{-4w} (-(dx^0)^2 + dw^2)\, .
 \ee
This is identical to \refb{ebc3} up to a relabelling of the
coordinate $\vp$ as $w$. Thus we see that the classical dynamics
of a D0-brane in the black hole and the usual linear dilaton
background are indeed identical. It is important to note that the
coordinate transformation that relates the two metrics involves
only a redefinition of the Liouville coordinate and does not
involve $x^0$. This is consistent with the fact that for both the
black hole and the linear dilaton background, the asymptotic time
coordinate $x^0$ is identified as the time coordinate of the
matrix model.

It is in fact easy to show that both \refb{ebc3} and \refb{ebc7} describe
a flat two dimensional metric\cite{0105038}. The easiest way to see this
is to begin
with the black hole background in the conformal gauge\cite{MSW,WB}:
 \be \label{ebc8}
\wt{ds}^2 = -e^{2\Phi_D} du dv , \qquad e^{-2\Phi_D} = a - 4uv\, .
 \ee
This gives:
 \be \label{ebc9}
\wt{ds}_{D0}^2 = e^{-2\Phi_D} \wt{ds}^2 = -du dv\, ,
 \ee
which is manifestly a flat metric. Thus in this coordinate system
the classical solutions correspond to D0-branes moving with
uniform velocity. Given such a solution we can convert the answer
to any other coordinate system. Note however that unlike the previous
transformation that related the black hole background to the
linear dilaton background, in this case the coordinate
transformation involves $x^0$ in a non-trivial fashion. Thus time
in the coordinate system in which the D0-brane sees a flat metric
does not coincide with the time coordinate of the matrix model.

Clearly there are many questions which remain to be answered, but
we hope that our results will provide a useful starting point for
a complete understanding of the two dimensional black hole in the
matrix model. Understanding this and other questions will
invariably lead to a deeper understanding of two dimensional
string theory and also possibly critical string theories.

\sectiono{Lessons for Critical String Theory} \label{scrit}

Although we have made heavy use of the closed string gauge
transformations, it should be noted that the final conclusions
about the existence of global symmetries and associated conserved
charges are for the open string field theory. For the D0-brane of
the two dimensional string theory the action of this open string
field theory can be taken to be the usual cubic action\cite{osft}
evaluated for the specific boundary CFT describing the D0-brane.
The open string spectrum in this theory is given by the product of
an arbitrary state in the $X$ and ghost CFT, and a state of the
Liouville theory built by the action of the Liouville Virasoro
generators on the SL(2,R) invariant vacuum\cite{0101152}. In
particular there is no open string state constructed over a
non-trivial primary of the Liouville sector. As a result the
relevant correlation functions needed for the construction of this
open string field theory involves arbitrary combination of $X$ and
ghost operators and the Liouville stress tensor. These can be
computed without any knowledge of the Liouville theory except its
total central charge which is $25$.

Now consider a different theory where we have replaced the
Liouville theory by an arbitrary $c=25$ CFT. Let us take an
arbitrary D-brane in this theory and denote the corresponding
boundary CFT by $\FF$. The open string spectrum on this D-brane
will of course involve excitations over non-trivial primaries of
$\FF$. However irrespective of what $c=25$ theory we have, or what
D-brane we are considering, there will always be a universal
subsector of open string states obtained by restricting ourselves
to only the $c=25$ Virasoro descendants of the SL(2,R) invariant
vacuum of $\FF$ and arbitrary excitations in the ghost and $X$
sector.\footnote{This is in fact the same universal theory that
was used in \cite{0005036,0207107} for the construction of the
tachyon lump solution and the rolling tachyon solution in open
string field theory.} The open string field theory of this
universal sector will be identical to the open string field theory
of the D0-brane of the two dimensional string theory since all the
relevant correlation functions are identical. Thus understanding
the classical dynamics of the D0-brane in the two dimensional
string theory gives us direct information about the classical
dynamics of this universal subsector of the open string field
theory on any D-brane in any background. In particular open string
field configurations in this universal subsector will be
characterized by the same conserved charges as in the two
dimensional string theory.

The above discussion has been given using the open string
viewpoint, but we could also try to directly analyze these
conservation laws for a general D-brane system in a generic string
theory following the procedure given in sections \ref{s3a} and
\ref{s3b}. For this we need to construct the analogs of the states
$|\phi_{j,m}(p)\ra$ in this more general string theory. Since in
section \ref{s3a} we have given an algorithm for constructing
$|\phi_{j,m}(p)\ra$ beginning with two sets of chiral states,
$|Y^{L}_{j,m}(p)\ra$ and $|\OO^L_{j,m}(p)\ra$ (and their
right-moving analogs), all we need to do is to give the
generalization of these states. This can be done as follows. Let
us assume that the $c=25$ theory has one free scalar field $Z$,
and that the $z$-direction is orthogonal to the D-brane under
consideration so that we have Dirichlet boundary condition on $Z$.
We now replace in eqs.\refb{exx8}, \refb{exx5} (and analogous
equations for right-handed states) $V^L_{2(1-j)}$ by \be
\label{ecr1} e^{2\sqrt{j^2-1}\, Z_L}\, , \ee and the Virasoro
generators of the Liouville field by the total Virasoro generators
of the $c=25$ theory. Formally $e^{2\sqrt{j^2-1}\, Z_L}$ is a
primary of dimension $1-j^2$, -- same as that of $V^L_{2(1-j)}$ in
the Liouville theory. This defines the analogs of
$|\OO^L_{j-1,m}(p)\ra$ and $| Y^L_{j,m}(p)\ra$ in this more
general string theory. We can now define $ |\eta^L_{(j),m}(p)\ra$,
$|\psi^L_{(j),m}(p)\ra$ as in eq.\refb{exx6} and \refb{exx9}, and
$|\Lambda_{j,m}(p)\ra$, $|\phi_{j,m}(p)\ra$ through eqs.
\refb{exx11}, \refb{exx13}. This in turn can be used to define the
conserved charges through eq. \refb{e20}. Note that due to the
presence of the operator $e^{2\sqrt{j^2-1}\, Z}$, the gauge
transformation parameter $|\Lambda_{j,m}(p)\ra$ grows
exponentially for large $Z$. However since the D-brane is
localized at a finite value of $Z$, the action of the gauge
transformation on the degrees of freedom on the D-brane is
expected to be finite and give rise to finite conserved charges.

There is however a subtlety in this analysis. In general the
states $|\OO^L_{j-1,m}\ra$ defined this way may not be annihilated
by $Q_B$ when we express the state in terms of $Z$ oscillators.
What is guaranteed however is that the result is a linear
combination of null states of the $c=25$ Virasoro algebra. As a
result the right hand sides of eq.\refb{exx6}, \refb{exx9},
\refb{exx12} now could contain additional terms which are linear
combinations of these null states. The conservation law \refb{e19}
still holds if the inner product of the boundary state $\la\BB|$
with these null states vanish. Typically null states are
orthogonal to regular primary and secondary states, but have
non-zero inner product with states which are neither primary
states nor can be regarded as a secondary state over another
primary. Thus as long as the boundary state $|\BB\ra$ does not
have such states, the inner product of $\la\BB|$ with these null
states will vanish and the conservation laws will hold.

In particular if the D-brane is such that there is separate
conservation of the world-sheet energy momentum tensor associated
with the $c=25$ theory and the $X$ CFT at the boundary of the
world-sheet, then we expect that the null states of the $c=25$
Virasoro algebra will have vanishing one point function on the
disk and hence its inner product with the boundary state will
vanish. The rolling tachyon solution on a D0-brane of the critical
string theory of course satisfies this condition, and hence
possesses conserved charges. Computation of these conserved
charges can be done following the analysis of section \ref{s3b}.
The final expression will be given by eq.\refb{e24} with
$~_{liouville}\la \BB| V_{2(1-j)}(0)|0\ra_{liouville}$ replaced by
the one point function of $e^{2\sqrt{j^2-1} Z(0)}$ on the unit
disk. Let us denote this by $\la e^{2\sqrt{j^2-1} Z(0)}\ra_D$.
This in turn is obtained by first computing $\la e^{ik Z(0)}
\ra_D$, and then analytically continuing the result to
$k=-2i\sqrt{j^2-1}$. Thus for example if the D-brane is localized
at $z=a$, then we have $\la e^{ik Z(0)} \ra_D = C e^{ika}$ for
some known constant $C$, and hence $\la e^{2\sqrt{j^2-1}
Z(0)}\ra_D=C e^{2\sqrt{j^2-1} \, a}$.

\bigskip

\noindent {\bf Acknowledgement}: I would like to thank A.~Dhar,
D.~Jatkar, D.~Gaiotto, D.~Ghoshal, S.~Govindarajan, S.~F.~Hassan,
J.~Karczmarek, H.~Liu, J.~Maldacena, G.~Mandal, S.~Minwalla,
L.~Rastelli, A.~Strominger, S.~Wadia and B.~Zwiebach for useful
discussions, and B.~Zwiebach for his comments on an earlier
version of the manuscript. A major part of the work was done
during my visit to the Center for Theoretical Physics at MIT as
Morningstar visiting professor. I would like to thank the members
of CTP for their hospitality. Some of the results of this paper
were presented at the Strings 2004 conference at Paris. I would
like to thank the organizers of the conference for hospitality and
for providing a stimulating environment during the conference. I
would also like to thank the University of Crete, CERN and the
Department of Physics of Stockholm University for their
hospitality during various stages of this work.

\appendix

\sectiono{Properties of $|\psi^L_{(j),m}\ra$ and
$|\eta^L_{(j),m}\ra$}\label{sappa}

In section \ref{s3a} we defined $|\eta^L_{(j),m}(p)\ra$ and
$|\psi^L_{(j),m}(p)\ra$ through the relations:
 \be \label{exx6a}
Q_B |\OO^L_{j-1,m}(p)\ra = (p-2m) |\eta^L_{(j),m}(p)\ra \, ,
 \ee
and
 \be \label{exx9a}
Q_B | Y^L_{j,m}(p)\ra = (p-2m) |\psi^L_{(j),m}(p)\ra\, ,
 \ee
where $|\OO^L_{j-1,m}(p)\ra$ and $|Y^L_{j,m}(p)\ra$ are chiral
states of conformal weight $({p^2\over 4}-m^2)$:
 \be \label{exx8a}
| Y^L_{j,m}(p)\ra = \PP^L_{j,m} \, e^{ip X_L(0)}|0\ra_X
\otimes V^L_{2(1-j)}(0) |0\ra_{liouville} \otimes
c_1|0\ra_{ghost} \, ,
 \ee
and
 \be \label{exx5a}
|\OO^L_{j-1,m}(p)\ra = \RR^L_{j-1,m} e^{i p X_L(0)} |0\ra_X
\otimes V^L_{2(1-j)}(0)|0\ra_{liouville} \otimes
c_1|0\ra_{ghost}
\, .
 \ee
 In this appendix we shall determine the SU(2) quantum numbers and
 other properties of  $|\eta^L_{(j),m}(p=2m)\ra$
and $|\psi^L_{(j),m}(p=2m)\ra$. We begin with the observation that
in a subspace with matter momentum $p$, $Q_B$ takes the form:
 \be \label{esp2}
Q_B = {1\over 4} \, p^2 \, c_0 + {1\over \sqrt 2} \,
p\, \sum_{n\ne 0} \, c_{-n} \alpha_n + \wh Q_B\, ,
 \ee
where $\wh Q_B$ does not depend on $p$. Knowing that $Q_B$
annihilates $| Y^L_{j,m}(p)\ra$ and $|\OO^L_{j-1,m}(p)\ra$ for
$p=2m$, and that the oscillator parts $\PP^L_{j,m}$,
$\RR^L_{j-1,m}$ involved in the construction of
$|Y^L_{j,m}(p)\ra$, $|\OO^L_{j-1,m}(p)\ra$ do not depend on $p$,
we can write:
 \ben \label{esp3}
Q_B | Y^L_{j,m}(p)\ra &=& \left({1\over 4} (p^2 - 4m^2) c_0 +
{1\over \sqrt 2} (p-2m) \sum_{n\ne 0} \, c_{-n} \alpha_n\right)|
Y^L_{j,m}(p)\ra \nonumber \\  Q_B | \OO^L_{j-1,m}(p)\ra &=&
\left({1\over 4} (p^2 - 4m^2) c_0 + {1\over \sqrt 2} (p-2m)
\sum_{n\ne 0} \, c_{-n} \alpha_n\right)| \OO^L_{j-1,m}(p)\ra \, .
 \een
Comparing this with \refb{exx6a}, \refb{exx9a} we get
 \ben \label{esp4}
|\psi^L_{(j),m}(p)\ra &=& {1\over 4} (p+2m) c_0 | Y^L_{j,m}(p)\ra
+ {1\over \sqrt 2} \, \sum_{n\ge 1} c_{-n} \alpha_n | Y^L_{j,m}(p)\ra
\nonumber \\
|\eta^L_{(j),m}(p)\ra &=& {1\over 4} (p+2m) c_0 | \OO^L_{j-1,m}(p)\ra
+ {1\over \sqrt 2} \, \sum_{n\ne 0} c_{-n} \alpha_n |
\OO^L_{j-1,m}(p)\ra
\, .
 \een
Note that the sum over $n$ in the first equation is restricted to
$n\ge 1$ since $| Y^L_{j,m}(p)\ra$ is annihilated by $c_n$ for
$n\ge 1$.

Let us now focus on the case $p=2m$. From \refb{esp4} it follows
that $|\eta^L_{(j),m}\ra\equiv|\eta^L_{(j),m}(p=2m)\ra$ and
$|\psi^L_{(j),m}\ra\equiv |\psi^L_{(j),m}(p=2m)\ra$ are given by
 \ben \label{eqnew}
 |\psi^L_{(j),m}\ra &=& m c_0 | Y^L_{j,m}\ra
+ {1\over \sqrt 2} \, \sum_{n\ge 1} c_{-n} \alpha_n | Y^L_{j,m}
\ra
\nonumber \\
|\eta^L_{(j),m}\ra &=& m c_0 | \OO^L_{j-1,m}\ra + {1\over \sqrt 2}
\, \sum_{n\ne 0} c_{-n} \alpha_n | \OO^L_{j-1,m}\ra \,.
 \een
 Clearly $|\eta^L_{(j),m}\ra$ and $|\psi^L_{(j),m}\ra$ must be
linear combination of states which are obtained by the action of
negative moded matter and Liouville Virasoro generators and
negative or zero moded ghost oscillators on states of the form:
 \be \label{exxnew}
|j',m\ra_L \otimes V^L_{2(1-j)}(0)|0\ra_{liouville} \otimes
c_1|0\ra_{ghost} \, ,
 \ee
for some $j'$. We want to analyze the possible values of $j'$
which could arise. The first terms in both equations on the right
hand side of \refb{eqnew} clearly have $j'=j$ and $j-1$
respectively since $c_0$ is an SU(2) singlet. On the other hand,
in the language of  SU(2) current algebra, $\alpha_{-n}$ is simply
the SU(2) current $J^3_{-n}$.  Thus the action of $\alpha_{-n}$ on
$| \OO^L_{j-1,m}\ra$ must be a linear combination of states of
spin lying between $j-2$ and $j$, and that on $| Y^L_{j,m}\ra$
must be a linear combination of states of spin $j$ and $j\pm 1$.
This leads us to the conclusion that $|\eta^L_{(j),m}\ra$ is built
on states of the form \refb{exxnew} with $(j-2)\le j'\le j$, and
$|\psi^L_{(j),m}\ra$ is built on states of the form \refb{exxnew}
with $(j-1)\le j'\le (j+1)$.

We can further restrict the structure of $|\psi^L_{(j),m}\ra$ and
$|\eta^L_{(j),m}\ra$ by analyzing their conformal weights. For
this we note that both $|\psi^L_{(j),m}\ra$ and
$|\eta^L_{(j),m}\ra$ have conformal weight 0. On the other hand
the state given in \refb{exxnew} has conformal weight $(j')^2 -
j^2$. Thus $|\eta^L_{(j),m}\ra$ and $|\psi^L_{(j),m}\ra$ can only
be linear combination of states built on states of the form
\refb{exxnew} with $j'\le j$. Furthermore, for $j'=j$, there
cannot be any non-zero mode oscillator acting on \refb{exxnew}.
Since $|\eta^L_{(j),m}\ra$ and $|\psi^L_{(j),m}\ra$ have ghost
numbers 1 and 2 respectively, it follows that the $j'=j$ state
appearing in the expression for $|\eta^L_{(j),m}\ra$ and
$|\psi^L_{(j),m}\ra$ must be proportional to $| Y^L_{j,m}\ra$ and
$c_0 | Y^L_{j,m}\ra$ respectively. It is known that the
coefficient of $| Y^L_{j,m}\ra$ appearing in the expression for
$|\eta^L_{(j),m}\ra$ is non-zero\cite{9108004}. We shall normalize
$|\OO^L_{j-1,m}\ra$ such that this coefficient is one. Thus we get
 \be \label{exi1a}
|\eta^L_{(j),m}\ra  = | Y^L_{j,m}\ra
+ |\wh \eta^L_{(j),m}\ra \, ,
 \ee
where $|\wh \eta^L_{(j),m}\ra$ is built on states $|j',m\ra$ with $j'=(j-1)$
or
$(j-2)$.

For $|\psi^L_{(j),m}\ra$, eq.\refb{eqnew}, together with the
constraints based on SU(2) symmetry, ghost number and conformal
weight, give
 \be \label{esp0a}
|\psi^L_{(j),m}\ra  = m \, c_0\, | Y^L_{j,m}\ra +
|\tau^L_{j-1,m}\ra
 \ee
where
 \be \label{espp1}
|\tau^L_{j-1,m}\ra = {1\over \sqrt 2} \, \sum_{n=1}^\infty c_{-n}
\alpha_n | Y^L_{j,m}\ra \, ,
 \ee
must be built by the action of X and Liouville Virasoro and ghost
oscillators on a state of the form \refb{exxnew} with $j'=(j-1)$.

\sectiono{Normalization of $\wh Q_{j,m}$} \label{appb}

In this appendix we shall compare the expression \refb{emore3} for
the charges $\wh Q_{j,m}$ with those found in
\cite{9108004,9109032,9210105,9209011} by comparing the symmetry
algebras in the continuum theory and the matrix model. In
particular \cite{9210105} constructed a set of symmetry generators
$\ov Q_{j,m}$ in the continuum theory at $\mu=0$ with Poisson
bracket: \be \label{ebb1}
 \{\ov Q_{j,m}, \ov Q_{j',m'}\} = 2\, (jm'-j'm)\, \,
 \ov Q_{j+j'-1,m+m'}\, .
 \ee
 Comparing this with the Poisson brackets computed using
 \refb{emore3} we see that it is consistent to make
 the assignment:
 \be \label{ebb2}
 \ov Q_{j,m} = -\left[(2j-1)! (j+m)! (j-m)!\right] \, \wh Q_{j,m}\, .
 \ee

 The charges $\ov Q_{j,m}$ found in \cite{9210105} were related to
 a set of symmetry transformation parameters
 $|\ov\Lambda_{j,m}\ra$ given by
 expressions similar to \refb{exx11} with $Y^{L,R}_{j,m}$,
 $\OO^{L,R}_{j,m}$ replaced by $\ov Y^{L,R}_{j,m}$ and $\ov
 \OO^{L,R}_{j,m}$ respectively:
  \be \label{ebb3}
 |\ov\Lambda_{j,m}\ra \sim {1\over 2} \left[ |\ov\OO^L_{j-1,m}\ra
\times | \ov Y^R_{j,m}\ra - | \ov Y^L_{j,m}\ra\times
|\ov\OO^R_{j-1,m}\ra \right] \, .
 \ee
 Here $\sim$ denotes equality up to an overall $j$ and $m$
 independent numerical factor.
 $\ov Y^{L,R}_{j,m}$ and $\ov \OO^{L,R}_{j,m}$ differ from
 $Y^{L,R}_{j,m}$ and $\OO^{L,R}_{j,m}$ used in this paper
 by normalization factors.
  The relative normalization between
 $Y^{L,R}_{j,m}$ and $\ov Y^{L,R}_{j,m}$ is given by (see eq.(2.6)
 of \cite{9210105}):
\be \label{ebb4}
 \ov Y^{L,R}_{j,m} = \left[(2j)! (j+m)! (j-m)!\right]^{1\over 2} \,
 Y^{L,R}_{j,m}\, .
 \ee
 We shall now try to find the relative normalization between
 $\OO^{L,R}_{j,m}$ and $\ov
 \OO^{L,R}_{j,m}$ so that we can find the relation between
 $|\ov\Lambda_{j,m}\ra$ and\footnote{As in section \ref{sadm} we shall
 assume that we are working with the tachyon potential $\mu
 e^{2\vp}$ from the beginning, and hence we can replace $\Lambda$
 in \refb{exx11}
 by $\wh\Lambda$. This way, while taking the $\mu\to 0$ limit we
 do not have to include any additional $\mu$ dependent multiplicative
 factor in the definition of the charges.}
 \be \label{ebb4a}
 |\wh\Lambda_{j,m}\ra = {1\over 2} \left[ |\OO^L_{j-1,m}\ra
\times | Y^R_{j,m}\ra - | Y^L_{j,m}\ra\times |\OO^R_{j-1,m}\ra
\right]
 \ee
 and compare this
 with \refb{ebb2}.

 $\ov \OO^{L,R}_{j,m}$ are normalized so that
 they satisfy the operator product algebra\cite{9210105}:
 \be \label{ebb5}
 \ov \OO^{L,R}_{j-1,m}\, \ov \OO^{L,R}_{j'-1,m'} =
 \ov \OO^{L,R}_{j+j'-2,m+m'} \, .
 \ee
 On the other hand according to \refb{eqnew}, \refb{exi1a}
 $\OO^{L,R}_{j-1,m}$ is normalized
 so that
 \be \label{ebb6}
 {1\over \sqrt 2}
 \, \sum_{n\ne 0} c_{-n} \alpha_n | \OO^L_{j-1,m}\ra = |Y^L_{j,m}\ra
 + \cdots\, , \qquad {1\over \sqrt 2}
 \, \sum_{n\ne 0} \bar c_{-n} \bar\alpha_n |
 \OO^R_{j-1,m}\ra = |Y^R_{j,m}\ra
 + \cdots\, ,
 \ee
 where $\cdots$ denote terms with SU(2) quantum number $(j-1,m)$
 and $(j-2,m)$.
By examining eq.\refb{ebb5} we see that the action of an SU(2)
rotation matrix on $\ov\OO_{j,m}$ cannot be given by the standard
unitary spin $j$ representation, since in that case the right hand
side of the equation will contain a factor of the Clebsch-Gordon
coefficient
 \ben \label{ebb7}
 C^{j-1,j'-1,j+j'-2}_{m, m', m+m'} &=& \bigg[{(2j-2)!\over (j-1+m)!(j-1-m)!}\,
 {(2j'-2)!\over (j'-1+m')!(j'-1-m')!}\nonumber \\
 && \, {(j+j'+m+m'-2)! (j+j'-2-m-m')!\over
 (2(j+j'-2))!} \bigg]^{1\over 2}\, .
 \een
Thus if we want the SU(2) transformations to be represented by
unitary matrices, we need to choose the basis of operators to be:
\be \label{ebb8}
 \left[{(2j-2)! \over (j-1+m)!(j-1-m)!}\right]^{1\over 2}\,
 \ov\OO^{L,R}_{j-1,m}\, .
 \ee
 On the other hand by examining \refb{ebb6} we see that the
 operators $\OO^{L,R}_{j-1,m}$ also do not transform by unitary
 matrices under an SU(2) rotation,
since in that case the right hand side of this equation
 will contain a factor of\footnote{Note that since $Y^{L,R}_{j,m}$
 are normalized, they transform by unitary matrices under an SU(2)
 rotation.}
  \be \label{ebb9}
  C^{j-1,1,j}_{m, 0, m} = \left[ (j+m) (j-m) \over j (2j-1)
  \right]^{1\over 2}\, .
  \ee
   The operators
 which do transform in a unitary representation of SU(2) are of
 the form:
 \be \label{ebb10}
 \left[{(j+m) (j-m) \over j (2j-1)}\right]^{1\over 2} \, \OO^{L,R}_{j-1,m}\, .
 \ee
 Since the operators given in \refb{ebb8} and \refb{ebb10}
 transform in identical representations of the SU(2) group, they
 must be related by an $m$-independent multiplicative constant.
 This gives:
  \ben \label{ebb12}
\ov\OO^L_{j-1,m} &=& F(j) \, \left[{(j-1+m)!(j-1-m)! \over (2j-2)!
} \,
{(j+m) (j-m) \over j (2j-1)}\right]^{1\over 2} \, \OO^L_{j-1,m} \nonumber \\
&=& {\sqrt 2} \, F(j) \, \left[{(j+m)! (j-m)! \over (2j)!
}\right]^{1\over 2} \, \OO^L_{j-1,m} \, ,
 \een
 for some function $F(j)$ of $j$.

 Substituting \refb{ebb4},
 \refb{ebb12} and their right handed counterpart
 into \refb{ebb3} we see that
  \be \label{ebb13}
  |\ov\Lambda_{j,m}\ra \sim {\sqrt 2}\, F(j) \, (j+m)! \,
  (j-m)!\,
  \, |\Lambda_{j,m}\ra\, .
 \ee
Comparing \refb{ebb13} with \refb{ebb2} we see that at least the
$m$ dependence of the relative normalization between $\ov Q_{j,m}$
and $\wh Q_{j,m}$ is compatible with that between
$|\ov\Lambda_{j,m}\ra$ and $|\wh\Lambda_{j,m}\ra$.

In order to compare the $j$ dependence of the relative
normalization of $Q_{j,m}$ and $\ov Q_{j,m}$ with that of
$\Lambda_{j,m}$ and $\ov\Lambda_{j,m}$, we can choose a convenient
value of $m$. We shall choose $m=j-1$. In this case from
\refb{ebb2} we get
 \be \label{ebb14}
 \ov Q_{j,j-1} = -\left[ (2j-1)!\right]^2 \, \wh Q_{j,j-1}\, .
 \ee
 Also \refb{ebb4} gives
 \be \label{ebb15}
 \ov Y^{L,R}_{j,j-1} = \left[ (2j)!(2j-1)!\right]^{1\over 2} \,
 Y^{L,R}_{j,j-1}\, .
 \ee
 To find the relation between $\ov\OO_{j-1,j-1}$ and $\OO_{j-1,j-1}$
 we note that \refb{ebb6} for $m=j-1$ gives
 \be \label{ebb16}
 {1\over \sqrt 2}
 \, \sum_{n\ne 0} c_{-n} \alpha_n | \OO^L_{j-1,j-1}\ra = |Y^L_{j,j-1}\ra
 + \cdots\, , \qquad {1\over \sqrt 2} \,
 \sum_{n\ne 0} \bar c_{-n} \bar\alpha_n
 | \OO^R_{j-1,j-1}\ra = |Y^R_{j,j-1}\ra
 + \cdots\, .
 \ee
 On the other hand an expression for $\ov\OO^{L,R}_{j,m}$ satisfying
 \refb{ebb5} was given in \cite{9209011} in terms of Schur
 polynomials. Using this expression and using the properties of
 the Schur polynomial one finds
 \be \label{ebb17}
{1\over \sqrt 2}
 \, \sum_{n\ne 0} c_{-n} \alpha_n | \ov\OO^L_{j-1,j-1}\ra =
 {1\over \sqrt 2} \,
 \left[ (2j)!(2j-1)!\right]^{1\over 2} \, |Y^L_{j,j-1}\ra
 + \cdots\, ,
 \ee
and a similar expression relating $|\ov\OO^R_{j-1,j-1}\ra$ to
$|Y^R_{j, j-1}\ra$. Comparing \refb{ebb16} and \refb{ebb17} we see
that
 \be \label{ebb18}
 |\ov\OO^{L,R}_{j-1,j-1}\ra =  {1\over \sqrt 2} \,
 \left[ (2j)!(2j-1)!\right]^{1\over 2} \, |\OO^{L,R}_{j-1,j-1}\ra \, .
 \ee
 Substituting \refb{ebb15} and \refb{ebb18} into
 eq.\refb{ebb3} we now get:
 \be \label{ebb19}
 |\ov\Lambda_{j,j-1}\ra \sim \left[ (2j)!(2j-1)!\right] \,
 |\wh\Lambda_{j,j-1}\ra\, .
 \ee

 Comparing \refb{ebb14} with \refb{ebb19} we see that the proportionality
 factor between $\ov Q_{j,m}$ and $\wh Q_{j,m}$ differs from that between $\ov
 \Lambda_{j,m}$ and $\wh \Lambda_{j,m}$ by a factor proportional to
 $j$. The precise meaning of this mismatch is not entirely clear
 to us. It appears as if the conserved charges constructed
 following the procedure of sections \ref{s2} and \ref{s3a} are
 related to the charges which generate the corresponding
 symmetries by a $j$-dependent normalization factor. Repeating a
 similar exercise for other states of this theory, {\it e.g.}
 the hole states might throw some light on this issue.


\begin{thebibliography}{99}

\bibitem{0304224}
J.~McGreevy and H.~Verlinde,
%``Strings from tachyons: The c = 1 matrix reloated,''
arXiv:hep-th/0304224.
%%CITATION = HEP-TH 0304224;%%

\bibitem{0305159}
I.~R.~Klebanov, J.~Maldacena and N.~Seiberg,
%``D-brane decay in two-dimensional string theory,''
JHEP {\bf 0307}, 045 (2003) [arXiv:hep-th/0305159].
%%CITATION = HEP-TH 0305159;%%

\bibitem{0305194}
J.~McGreevy, J.~Teschner and H.~Verlinde,
%``Classical and quantum D-branes in 2D string theory,''
arXiv:hep-th/0305194.
%%CITATION = HEP-TH 0305194;%%

\bibitem{0307083}
T.~Takayanagi and N.~Toumbas,
%``A matrix model dual of type 0B string theory in two dimensions,''
JHEP {\bf 0307}, 064 (2003) [arXiv:hep-th/0307083].
%%CITATION = HEP-TH 0307083;%%

\bibitem{0307195}
M.~R.~Douglas, I.~R.~Klebanov, D.~Kutasov, J.~Maldacena,
E.~Martinec and N.~Seiberg,
%``A new hat for the c = 1 matrix model,''
arXiv:hep-th/0307195.
%%CITATION = HEP-TH 0307195;%%

\bibitem{0308068}
A.~Sen,
%``Open-closed duality: Lessons from matrix model,''
Mod.\ Phys.\ Lett.\ A {\bf 19}, 841 (2004) [arXiv:hep-th/0308068].
%%CITATION = HEP-TH 0308068;%%

\bibitem{0305011}
A.~Sen,
%``Open and closed strings from unstable D-branes,''
Phys.\ Rev.\ D {\bf 68}, 106003 (2003) [arXiv:hep-th/0305011].
%%CITATION = HEP-TH 0305011;%%

\bibitem{0306137}
A.~Sen,
%``Open-closed duality at tree level,''
Phys.\ Rev.\ Lett.\  {\bf 91}, 181601 (2003)
[arXiv:hep-th/0306137].
%%CITATION = HEP-TH 0306137;%%

\bibitem{0312135}
J.~de Boer, A.~Sinkovics, E.~Verlinde and J.~T.~Yee,
%``String interactions in c = 1 matrix model,''
JHEP {\bf 0403}, 023 (2004) [arXiv:hep-th/0312135].
%%CITATION = HEP-TH 0312135;%%

\bibitem{0312163}
J.~Ambjorn and R.~A.~Janik,
%``The decay of quantum D-branes,''
Phys.\ Lett.\ B {\bf 584}, 155 (2004) [arXiv:hep-th/0312163].
%%CITATION = HEP-TH 0312163;%%

\bibitem{0312192}
G.~Mandal and S.~R.~Wadia,
%``Rolling tachyon solution of two-dimensional string theory,''
arXiv:hep-th/0312192.
%%CITATION = HEP-TH 0312192;%%

\bibitem{0312196}
D.~Gaiotto and L.~Rastelli,
%``A paradigm of open/closed duality: Liouville D-branes and the Kontsevich
%model,''
arXiv:hep-th/0312196.
%%CITATION = HEP-TH 0312196;%%

\bibitem{0406106}
D.~Ghoshal, S.~Mukhi and S.~Murthy,
%``Liouville D-branes in two-dimensional strings and open string field theory,''
arXiv:hep-th/0406106.
%%CITATION = HEP-TH 0406106;%%


\bibitem{DAVID}
F.~David,
%``Conformal Field Theories Coupled To 2-D Gravity In The Conformal
%Gauge,''
Mod.\ Phys.\ Lett.\ A {\bf 3}, 1651 (1988).
%%CITATION = MPLAE,A3,1651;%%

\bibitem{DK}
J.~Distler and H.~Kawai,
%``Conformal Field Theory And 2-D Quantum Gravity Or Who's Afraid Of
%Joseph
%Liouville?,''
Nucl.\ Phys.\ B {\bf 321}, 509 (1989).
%%CITATION = NUPHA,B321,509;%%

\bibitem{KPZ}
V.~G.~Knizhnik, A.~M.~Polyakov and A.~B.~Zamolodchikov,
%``Fractal Structure Of 2d-Quantum Gravity,''
Mod.\ Phys.\ Lett.\ A {\bf 3}, 819 (1988).
%%CITATION = MPLAE,A3,819;%%

\bibitem{GROMIL}
D.~J.~Gross and N.~Miljkovic,
%``A Nonperturbative Solution Of D = 1 String Theory,''
Phys.\ Lett.\ B {\bf 238}, 217 (1990).
%%CITATION = PHLTA,B238,217;%%

\bibitem{BKZ}
E.~Brezin, V.~A.~Kazakov and A.~B.~Zamolodchikov,
%``Scaling Violation In A Field Theory Of Closed Strings In One Physical
%Dimension,''
Nucl.\ Phys.\ B {\bf 338}, 673 (1990).
%%CITATION = NUPHA,B338,673;%%

\bibitem{GINZIN}
P.~Ginsparg and J.~Zinn-Justin,
%``2-D Gravity + 1-D Matter,''
Phys.\ Lett.\ B {\bf 240}, 333 (1990).
%%CITATION = PHLTA,B240,333;%%

\bibitem{DASJEV}
S.~R.~Das and A.~Jevicki,
%``String Field Theory And Physical Interpretation Of D = 1 Strings,''
Mod.\ Phys.\ Lett.\ A {\bf 5}, 1639 (1990).
%%CITATION = MPLAE,A5,1639;%%

\bibitem{SENWAD}
A.~M.~Sengupta and S.~R.~Wadia,
%``Excitations And Interactions In D = 1 String Theory,''
Int.\ J.\ Mod.\ Phys.\ A {\bf 6}, 1961 (1991).
%%CITATION = IMPAE,A6,1961;%%

\bibitem{GROKLE}
D.~J.~Gross and I.~R.~Klebanov,
%``Fermionic String Field Theory Of C = 1 Two-Dimensional Quantum
%Gravity,''
Nucl.\ Phys.\ B {\bf 352}, 671 (1991).
%%CITATION = NUPHA,B352,671;%%

\bibitem{0101152}
A.~B.~Zamolodchikov and A.~B.~Zamolodchikov,
%``Liouville field theory on a pseudosphere,''
arXiv:hep-th/0101152.
%%CITATION = HEP-TH 0101152;%%

\bibitem{0203211}
A.~Sen,
%``Rolling tachyon,''
JHEP {\bf 0204}, 048 (2002) [arXiv:hep-th/0203211].
%%CITATION = HEP-TH 0203211;%%

\bibitem{0203265}
A.~Sen,
%``Tachyon matter,''
JHEP {\bf 0207}, 065 (2002) [arXiv:hep-th/0203265].
%%CITATION = HEP-TH 0203265;%%

\bibitem{0303139}
N.~Lambert, H.~Liu and J.~Maldacena,
%``Closed strings from decaying D-branes,''
arXiv:hep-th/0303139.
%%CITATION = HEP-TH 0303139;%%

\bibitem{0304192}
D.~Gaiotto, N.~Itzhaki and L.~Rastelli,
%``Closed strings as imaginary D-branes,''
Nucl.\ Phys.\ B {\bf 688}, 70 (2004) [arXiv:hep-th/0304192].
%%CITATION = HEP-TH 0304192;%%

\bibitem{0306132}
J.~L.~Karczmarek, H.~Liu, J.~Maldacena and A.~Strominger,
%``UV finite brane decay,''
JHEP {\bf 0311}, 042 (2003) [arXiv:hep-th/0306132].
%%CITATION = HEP-TH 0306132;%%

\bibitem{0307221}
D.~Gaiotto, N.~Itzhaki and L.~Rastelli,
%``On the BCFT description of holes in the c = 1 matrix model,''
Phys.\ Lett.\ B {\bf 575}, 111 (2003) [arXiv:hep-th/0307221].
%%CITATION = HEP-TH 0307221;%%

\bibitem{MSW}
G.~Mandal, A.~M.~Sengupta and S.~R.~Wadia,
%``Classical solutions of two-dimensional string theory,''
Mod.\ Phys.\ Lett.\ A {\bf 6}, 1685 (1991).
%%CITATION = MPLAE,A6,1685;%%

\bibitem{WB}
E.~Witten,
%``On string theory and black holes,''
Phys.\ Rev.\ D {\bf 44}, 314 (1991).
%%CITATION = PHRVA,D44,314;%%

\bibitem{0101011}
V.~Kazakov, I.~K.~Kostov and D.~Kutasov,
%``A matrix model for the two-dimensional black hole,''
Nucl.\ Phys.\ B {\bf 622}, 141 (2002) [arXiv:hep-th/0101011].
%%CITATION = HEP-TH 0101011;%%

\bibitem{0311177}
S.~Dasgupta and T.~Dasgupta,
%``Nonsinglet sector of c = 1 matrix model and 2D black hole,''
arXiv:hep-th/0311177.
%%CITATION = HEP-TH 0311177;%%

\bibitem{MOORESEI}
G.~W.~Moore and N.~Seiberg,
%``From loops to fields in 2-D quantum gravity,''
Int.\ J.\ Mod.\ Phys.\ A {\bf 7} (1992) 2601.
%%CITATION = IMPAE,A7,2601;%%

\bibitem{MINIC}
D.~Minic, J.~Polchinski and Z.~Yang,
%``Translation Invariant Backgrounds In (1+1)-Dimensional String Theory,''
Nucl.\ Phys.\ B {\bf 369}, 324 (1992).
%%CITATION = NUPHA,B369,324;%%

\bibitem{9108004}
E.~Witten,
%``Ground ring of two-dimensional string theory,''
Nucl.\ Phys.\ B {\bf 373}, 187 (1992) [arXiv:hep-th/9108004].
%%CITATION = HEP-TH 9108004;%%

\bibitem{9109032}
I.~R.~Klebanov and A.~M.~Polyakov,
%``Interaction of discrete states in two-dimensional string theory,''
Mod.\ Phys.\ Lett.\ A {\bf 6}, 3273 (1991) [arXiv:hep-th/9109032].
%%CITATION = HEP-TH 9109032;%%

\bibitem{9110021}
S.~R.~Das, A.~Dhar, G.~Mandal and S.~R.~Wadia,
%``Gauge theory formulation of the C = 1 matrix model: Symmetries and
%discrete
%states,''
Int.\ J.\ Mod.\ Phys.\ A {\bf 7}, 5165 (1992)
[arXiv:hep-th/9110021].
%%CITATION = HEP-TH 9110021;%%

\bibitem{9209036}
J.~Avan and A.~Jevicki,
%``Interacting theory of collective and topological fields in
%two-dimensions,''
Nucl.\ Phys.\ B {\bf 397}, 672 (1993) [arXiv:hep-th/9209036].
%%CITATION = HEP-TH 9209036;%%

\bibitem{9210105}
I.~R.~Klebanov and A.~Pasquinucci,
%``Infinite asymmetry and Ward identities in two-dimensional string
%theory,''
arXiv:hep-th/9210105.
%%CITATION = HEP-TH 9210105;%%

\bibitem{9302106}
A.~Jevicki,
%``Fields and symmetries of 2-D strings,''
arXiv:hep-th/9302106.
%%CITATION = HEP-TH 9302106;%%

\bibitem{9201056}
E.~Witten and B.~Zwiebach,
%``Algebraic structures and differential geometry in $2-D$ string
%theory,''
Nucl.\ Phys.\ B {\bf 377}, 55 (1992) [arXiv:hep-th/9201056].
%%CITATION = HEP-TH 9201056;%%

\bibitem{9507041}
A.~Dhar, G.~Mandal and S.~R.~Wadia,
%``Discrete state moduli of string theory from the C=1 matrix model,''
Nucl.\ Phys.\ B {\bf 454}, 541 (1995) [arXiv:hep-th/9507041].
%%CITATION = HEP-TH 9507041;%%

\bibitem{0402157}
A.~Sen, arXiv:hep-th/0402157.
%%CITATION = HEP-TH 0402157;%%

\bibitem{9705241}
B.~Zwiebach,
%``Oriented open-closed string theory revisited,''
Annals Phys.\  {\bf 267}, 193 (1998) [arXiv:hep-th/9705241], and
references therein.
%%CITATION = HEP-TH 9705241;%%

\bibitem{LIAN}
B.~H.~Lian and G.~J.~Zuckerman, Phys.\ Lett.\ B {\bf 254}, 417
(1991);
%%CITATION = PHLTA,B254,417;%%
%``2-D Gravity With C = 1 Matter,''
Phys.\ Lett.\ B {\bf 266}, 21 (1991).
%%CITATION = PHLTA,B266,21;%%

\bibitem{MMS1}
S.~Mukherji, S.~Mukhi and A.~Sen,
%``Null vectors and extra states in c = 1 string theory,''
Phys.\ Lett.\ B {\bf 266}, 337 (1991).
%%CITATION = PHLTA,B266,337;%%

\bibitem{BMP}
P.~Bouwknegt, J.~G.~McCarthy and K.~Pilch,
%``BRST analysis of physical states for 2-D gravity coupled to c <= 1
%matter,''
Commun.\ Math.\ Phys.\  {\bf 145}, 541 (1992).
%%CITATION = CMPHA,145,541;%%

\bibitem{MMS}
S.~Mukherji, S.~Mukhi and A.~Sen,
%``Black hole solution and its infinite parameter generalizations in c = 1
%string field theory,''
Phys.\ Lett.\ B {\bf 275}, 39 (1992).
%%CITATION = PHLTA,B275,39;%%

\bibitem{osft}
E.~Witten,
%``Noncommutative Geometry And String Field Theory,''
Nucl.\ Phys.\ B {\bf 268}, 253 (1986).
%%CITATION = NUPHA,B268,253;%%

\bibitem{0403200}
A.~Sen,
%``Energy momentum tensor and marginal deformations in open string field
%theory,''
arXiv:hep-th/0403200.
%%CITATION = HEP-TH 0403200;%%

\bibitem{0309074}
T.~Asakawa, S.~Kobayashi and S.~Matsuura,
%``Closed string field theory with dynamical D-brane,''
JHEP {\bf 0310}, 023 (2003) [arXiv:hep-th/0309074].
%%CITATION = HEP-TH 0309074;%%

\bibitem{polch}
J.~Polchinski,
%``Critical Behavior Of Random Surfaces In One-Dimension,''
Nucl.\ Phys.\ B {\bf 346}, 253 (1990).
%%CITATION = NUPHA,B346,253;%%

\bibitem{DDW}
S.~R.~Das, A.~Dhar and S.~R.~Wadia,
%``Critical Behavior In Two-Dimensional Quantum Gravity And Equations Of
%Motion
%Of The String,''
Mod.\ Phys.\ Lett.\ A {\bf 5}, 799 (1990).
%%CITATION = MPLAE,A5,799;%%

\bibitem{DVV}
R.~Dijkgraaf, E.~Verlinde and H.~Verlinde,
%``C = 1 Conformal Field Theories On Riemann Surfaces,''
Commun.\ Math.\ Phys.\  {\bf 115}, 649 (1988).
%%CITATION = CMPHA,115,649;%%

\bibitem{goli}
M.~Goulian and M.~Li,
%``Correlation Functions In Liouville Theory,''
Phys.\ Rev.\ Lett.\  {\bf 66}, 2051 (1991).
%%CITATION = PRLTA,66,2051;%%

\bibitem{9206053}
H.~Dorn and H.~J.~Otto,
%``On correlation functions for noncritical strings with c <= 1 d >= 1,''
Phys.\ Lett.\ B {\bf 291}, 39 (1992) [arXiv:hep-th/9206053].
%%CITATION = HEP-TH 9206053;%%

\bibitem{9403141}
H.~Dorn and H.~J.~Otto,
%``Two and three point functions in Liouville theory,''
Nucl.\ Phys.\ B {\bf 429}, 375 (1994) [arXiv:hep-th/9403141].
%%CITATION = HEP-TH 9403141;%%

\bibitem{9506136}
A.~B.~Zamolodchikov and A.~B.~Zamolodchikov,
%``Structure constants and conformal bootstrap in Liouville field
%theory,''
Nucl.\ Phys.\ B {\bf 477}, 577 (1996) [arXiv:hep-th/9506136].
%%CITATION = HEP-TH 9506136;%%

\bibitem{0104158}
J.~Teschner,
%``Liouville theory revisited,''
Class.\ Quant.\ Grav.\  {\bf 18}, R153 (2001)
[arXiv:hep-th/0104158].
%%CITATION = HEP-TH 0104158;%%

\bibitem{IMM}
C.~Imbimbo, S.~Mahapatra and S.~Mukhi,
%``Construction of physical states of nontrivial ghost number in c < 1
%string
%theory,''
Nucl.\ Phys.\ B {\bf 375}, 399 (1992).
%%CITATION = NUPHA,B375,399;%%

\bibitem{9209011}
Y.~S.~Wu and C.~J.~Zhu,
%``The Complete structure of the cohomology ring and associated symmetries
%in D
%= 2 string theory,''
Nucl.\ Phys.\ B {\bf 404}, 245 (1993) [arXiv:hep-th/9209011].
%%CITATION = HEP-TH 9209011;%%

\bibitem{ishibashi}
N.~Ishibashi,
%``The Boundary And Crosscap States In Conformal Field Theories,''
Mod.\ Phys.\ Lett.\ A {\bf 4}, 251 (1989).
%%CITATION = MPLAE,A4,251;%%

\bibitem{9402113}
C.~G.~.~Callan, I.~R.~Klebanov, A.~W.~W.~Ludwig and
J.~M.~Maldacena,
%``Exact solution of a boundary conformal field theory,''
Nucl.\ Phys.\ B {\bf 422}, 417 (1994) [arXiv:hep-th/9402113].
%%CITATION = HEP-TH 9402113;%%

\bibitem{9811237}
 A.~Recknagel and V.~Schomerus,
%``Boundary deformation theory and moduli spaces of D-branes,''
Nucl.\ Phys.\ B {\bf 545}, 233 (1999) [arXiv:hep-th/9811237].
%%CITATION = HEP-TH 9811237;%%

\bibitem{0208196}
T.~Okuda and S.~Sugimoto,
%``Coupling of rolling tachyon to closed strings,''
Nucl.\ Phys.\ B {\bf 647}, 101 (2002) [arXiv:hep-th/0208196].
%%CITATION = HEP-TH 0208196;%%

\bibitem{0212248}
F.~Larsen, A.~Naqvi and S.~Terashima,
%``Rolling tachyons and decaying branes,''
JHEP {\bf 0302}, 039 (2003) [arXiv:hep-th/0212248].
%%CITATION = HEP-TH 0212248;%%

\bibitem{0305177}
N.~R.~Constable and F.~Larsen,
%``The rolling tachyon as a matrix model,''
JHEP {\bf 0306}, 017 (2003) [arXiv:hep-th/0305177].
%%CITATION = HEP-TH 0305177;%%

\bibitem{0301038}
M.~Gutperle and A.~Strominger,
%``Timelike boundary Liouville theory,''
Phys.\ Rev.\ D {\bf 67}, 126002 (2003) [arXiv:hep-th/0301038].
%%CITATION = HEP-TH 0301038;%%

\bibitem{0308172}
K.~Okuyama,
%``Comments on half S-branes,''
JHEP {\bf 0309}, 053 (2003) [arXiv:hep-th/0308172].
%%CITATION = HEP-TH 0308172;%%

\bibitem{seiberg}
N.~Seiberg,
%``Notes On Quantum Liouville Theory And Quantum Gravity,''
Prog.\ Theor.\ Phys.\ Suppl.\  {\bf 102}, 319 (1990).
%%CITATION = PTPSA,102,319;%%

\bibitem{0406030}
D.~Kutasov, K.~Okuyama, J.~Park, N.~Seiberg, D.~Shih
[arXiv:hep-th/0406030].
%%CITATION = HEP-TH 0406030;%%

\bibitem{0401067}
S.~R.~Das,
%``D branes in 2d string theory and classical limits,''
arXiv:hep-th/0401067.
%%CITATION = HEP-TH 0401067;%%


\bibitem{cardy}
J.~L.~Cardy,
%``Boundary Conditions, Fusion Rules And The Verlinde Formula,''
Nucl.\ Phys.\ B {\bf 324}, 581 (1989).
%%CITATION = NUPHA,B324,581;%%

\bibitem{0405058}
D.~Kutasov,
%``D-brane dynamics near NS5-branes,''
arXiv:hep-th/0405058.
%%CITATION = HEP-TH 0405058;%%

\bibitem{0204143}
A.~Sen,
%``Field theory of tachyon matter,''
Mod.\ Phys.\ Lett.\ A {\bf 17}, 1797 (2002)
[arXiv:hep-th/0204143].
%%CITATION = HEP-TH 0204143;%%

\bibitem{0406173}
Y.~Nakayama, Y.~Sugawara and H.~Takayanagi,
%``Boundary states for the rolling D-branes in NS5 background,''
arXiv:hep-th/0406173.
%%CITATION = HEP-TH 0406173;%%

\bibitem{0310024}
S.~Ribault and V.~Schomerus,
%``Branes in the 2-D black hole,''
JHEP {\bf 0402}, 019 (2004) [arXiv:hep-th/0310024].
%%CITATION = HEP-TH 0310024;%%

\bibitem{0312168}
S.~L.~Lukyanov, E.~S.~Vitchev and A.~B.~Zamolodchikov,
%``Integrable model of boundary interaction: The paperclip,''
Nucl.\ Phys.\ B {\bf 683}, 423 (2004) [arXiv:hep-th/0312168].
%%CITATION = HEP-TH 0312168;%%

\bibitem{0001012}
V.~Fateev, A.~B.~Zamolodchikov and A.~B.~Zamolodchikov,
%``Boundary Liouville field theory. I: Boundary state and boundary  two-point
%function,''
arXiv:hep-th/0001012.
%%CITATION = HEP-TH 0001012;%%

\bibitem{0009138}
J.~Teschner,
%``Remarks on Liouville theory with boundary,''
arXiv:hep-th/0009138.
%%CITATION = HEP-TH 0009138;%%


\bibitem{9210107}
S.~R.~Das,
%``Matrix models and black holes,''
Mod.\ Phys.\ Lett.\ A {\bf 8}, 69 (1993) [arXiv:hep-th/9210107].
%%CITATION = HEP-TH 9210107;%%

\bibitem{marsha}
E.~J.~Martinec and S.~L.~Shatashvili,
%``Black hole physics and Liouville theory,''
Nucl.\ Phys.\ B {\bf 368}, 338 (1992).
%%CITATION = NUPHA,B368,338;%%

\bibitem{9210120}
A.~Dhar, G.~Mandal and S.~R.~Wadia,
%``Stringy quantum effects in two-dimensional black hole,''
Mod.\ Phys.\ Lett.\ A {\bf 7}, 3703 (1992) [arXiv:hep-th/9210120].
%%CITATION = HEP-TH 9210120;%%

\bibitem{9303116}
S.~R.~Das,
%``Matrix Models And Nonperturbative String Propagation In
%Two-Dimensional Black
%Hole Backgrounds,''
Mod.\ Phys.\ Lett.\ A {\bf 8}, 1331 (1993) [arXiv:hep-th/9303116].
%%CITATION = HEP-TH 9303116;%%

\bibitem{9304072}
A.~Dhar, G.~Mandal and S.~R.~Wadia,
%``Wave propagation in stringy black hole,''
Mod.\ Phys.\ Lett.\ A {\bf 8}, 1701 (1993) [arXiv:hep-th/9304072].
%%CITATION = HEP-TH 9304072;%%

\bibitem{9305109}
A.~Jevicki and T.~Yoneya,
%``A Deformed matrix model and the black hole background in two-dimensional
%string theory,''
Nucl.\ Phys.\ B {\bf 411}, 64 (1994) [arXiv:hep-th/9305109].
%%CITATION = HEP-TH 9305109;%%


\bibitem{mald}
J.~Maldacena, Talk at Strings 2004.

\bibitem{asparis}
A.~Sen, Talk at Strings 2004, Paris
(http://strings04.lpthe.jussieu.fr/talks/Sen.pdf).

\bibitem{0407136}
E.~Martinec and K.~Okuyama, arXiv:hep-th/0407136.
%%CITATION = HEP-TH 0407136;%%

\bibitem{0105038}
J.~M.~Maldacena, G.~W.~Moore and N.~Seiberg,
%``Geometrical interpretation of D-branes in gauged WZW models,''
JHEP {\bf 0107}, 046 (2001) [arXiv:hep-th/0105038].
%%CITATION = HEP-TH 0105038;%%

\bibitem{0406017}
A.~Fotopoulos, V.~Niarchos and N.~Prezas,
%``D-branes and extended characters in SL(2,R)/U(1),''
arXiv:hep-th/0406017.
%%CITATION = HEP-TH 0406017;%%

\bibitem{yogendran}
K.~P.~Yogendran, to appear.

\bibitem{0005036}
N.~Moeller, A.~Sen and B.~Zwiebach,
%``D-branes as tachyon lumps in string field theory,''
JHEP {\bf 0008}, 039 (2000) [arXiv:hep-th/0005036].
%%CITATION = HEP-TH 0005036;%%

\bibitem{0207107}
N.~Moeller and B.~Zwiebach,
%``Dynamics with infinitely many time derivatives and rolling tachyons,''
JHEP {\bf 0210}, 034 (2002) [arXiv:hep-th/0207107].
%%CITATION = HEP-TH 0207107;%%


\end{thebibliography}
\end{document}